\documentclass[reprint,amsmath,amssymb,aps,table]{revtex4-2}
\usepackage[latin9]{inputenc}
\usepackage{color}
\usepackage{prettyref}
\usepackage{float}
\usepackage{mathrsfs}
\usepackage{amsmath}
\usepackage{graphicx}

\makeatletter
\usepackage{xcolor}                       
\usepackage[colorlinks=true,               
            linkcolor=blue,                
            citecolor=blue,                
            urlcolor=blue]{hyperref}      
\usepackage{cleveref}
\usepackage{algorithm}
\usepackage{algpseudocode}

\usepackage{braket}
\usepackage{tikz}
\usetikzlibrary{calc,tikzmark}
\usepackage{adjustbox}

\usepackage{mathrsfs}

\usepackage{dcolumn}
\usepackage{bm}

\usepackage{tabularx}
\usepackage{makecell}
\usepackage{mathrsfs} 
\usepackage{makecell}
\usepackage{braket}
\usepackage{subfig}
\usepackage{subcaption}
\usepackage{tabularx}
\usepackage{booktabs}
\usepackage{siunitx}

\crefname{figure}{Figure}{Figures}
\crefname{table}{Table}{Tables}
\crefname{section}{Section}{Sections}
\crefname{subsection}{Section}{Sections}
\crefname{chapter}{Chapter}{Chapters}  
\crefname{appendix}{Appendix}{Appendices}

\Crefname{figure}{Figure}{Figures}
\Crefname{table}{Table}{Tables}
\Crefname{section}{Section}{Sections}
\Crefname{subsection}{Section}{Sections}
\Crefname{chapter}{Chapter}{Chapters}
\Crefname{appendix}{Appendix}{Appendices}

\makeatother
\usepackage{babel}
\begin{document}
\title{Reduction of interaction order in hard combinatorial optimization
via conditionally independent degrees of freedom}
\author{Alexandru Ciobanu\textsuperscript{1,2,3}}
\email{a.ciobanu@fz-juelich.de}
\author{David Dahmen\textsuperscript{1}}
\author{John Paul Strachan\textsuperscript{2,5}}
\author{Moritz Helias\textsuperscript{1,4}}

\affiliation{\textsuperscript{1}Institute for Advanced Simulation (IAS6), Computational and Systems Neuroscience, J\"ulich Research Centre, J\"ulich, Germany}
\affiliation{\textsuperscript{2}Neuromorphic Compute Nodes (PGI-14), Peter Gr\"unberg Institut (PGI), J\"ulich Research Centre, J\"ulich, Germany}
\affiliation{\textsuperscript{3}RWTH Aachen University, Aachen, Germany}
\affiliation{\textsuperscript{4}Department of Physics, Faculty 1, RWTH Aachen University, Aachen, Germany}
\affiliation{\textsuperscript{5}Forschungszentrum J\"ulich GmbH, J\"ulich, Germany}

\begin{abstract}

Combinatorial optimization problems have a broad range of applications
and map to physical systems with complex dynamics. Among them, the
3-SAT problem is prominent due to its NP-complete nature. In physics
terms, its solution corresponds to finding the ground state of a disordered
Ising spin Hamiltonian with third-order, or tensor, interactions.
The large growth of the number of third-order interactions with number
of variables poses technical difficulties for the physical implementation
of minimizers. Therefore, researchers have proposed quadratization
techniques which reduce the order of the system, however, at the cost
of including additional degrees of freedom. Their inclusion induces
a drastic slow down in the minimization, which makes such procedures
technically infeasible for large problems. In this work, we take a
physics approach by employing the renormalization group to create
a pairwise interacting system from the original third-order system
while preserving the free energy. Our procedure utilizes additional
degrees of freedom that exhibit an independent dynamics provided the
original degrees of freedom are fixed. A step-wise trace of the extra
variables while running the minimization is therefore theoretically
manageable, yielding a state-dependent effective interaction. We use
the effective interaction to reconstruct the original third-order
energy spectrum, as this yields equal scaling of computations-to-ground-state
compared to the original tensor formulation. Here, the original degrees
of freedom interact with a subsystem that appears to be in a superposition
of an exponentially large number of states. In the zero-temperature
limit, the superposition concentrates on one state. Our spectrum-engineering
techniques reveal new routes toward the ground state of disordered
Ising systems, through Markov chains, and allow for efficient technological
implementations.

\end{abstract}
\maketitle

\section{Introduction}

The k-SAT problem class finds applications as diverse as AI and robotics
\cite{Kautz1992,Smith2008}, circuit verification \cite{Biere2003,Moskewicz2001},
air traffic control \cite{Stojadinovic2014}, operations research
\cite{doi:10.1287/opre.45.2.288}, genetics \cite{Aceto2004}, protein
folding \cite{Jiang2002,Dilkina2007}, programming \cite{AnacondaCondaPerformance2025},
cryptography \cite{RAMAMOORTHY20232539,Nohl2008,Barkan2003}, the
blockchain \cite{cryptoeprint:2006/254} and mathematics \cite{Heule2016},
among many others \cite{anderson1980spinglass6}. Due to their NP-completeness
\cite{Aho1974}, these problems can be converted to the 3-SAT problem
which seeks, among all binary words of length $N$, those instances
which satisfy the complete set of Boolean expressions in which triples
of variables are combined in a logical disjunction. The solution that
satisfies simultaneously all Boolean expressions natively maps to
finding the ground state of a third-order Ising Hamiltonian \cite{PhysRevE.56.1357}.

Implementing third-order interactions poses a technical problem if
one wishes to employ specialized hardware known as Ising Machines.
These constitute efficient machines that implement networks of interacting
binary spin variables \cite{Hopfield1985,Mohseni2022,ErcseyRavasz2011,inbook,Byrnes2013,Chen2025,https://doi.org/10.1049/ell2.70236,Cen2022,Aadit2022},
but utilize mostly second-order interactions between variables \cite{Bybee2023}.
Higher-order Ising models have received less attention because the
number of possible interactions grows exponentially with the interaction
order \cite{10.1063/1.36246}. Therefore, a pre-processing step is
usually applied, to convert a higher-order Hamiltonian (known as PUBO
-- Polynomial Unconstrained Binary Optimization, or general interactions
of order higher than two) to QUBO (Quadratic Unconstrained Binary
Optimization, or general pairwise interactions). Interest in quadratic
Hamiltonians (QUBO) is also due to the developments in quantum computing,
as QUBO serves as an input format for quantum annealers and the quantum
approximate optimization algorithm on gate-based quantum computers
\cite{dwave2023guide,PRXQuantum.4.030335,Jiang2018,Wang2020,9286230}.
In addition, quadratic energy functions are ubiquitous in physical
systems \cite{Lin2017}, amenable to theoretical techniques and offer
an efficient physical implementation.

However, the existent quadratization techniques e.g.\cite{rosenberg1972breves,PhysRevA.77.052331,dattani2019quadratizationdiscreteoptimizationquantum,Anthony2017,PhysRevE.110.045308},
and variants thereof, admit several shortcomings today \cite{10558658}:

1) Most importantly, they introduce additional variables that not
only largely increase the search space, but also create a misalignment
between the original cost function and the quadratized version, i.e.
the ground states of the PUBO system correspond to configurations
in QUBO that do not lie at the minimum of the respective energy.

2) The additional variables impose ``ruggedness'' in the energy
landscape \cite{KAUFFMAN198711,PhysRevLett.35.1792,PhysRevE.110.045308};
the additional configurations induce local fields that effectively
guide the state towards local minima, hindering a local solver's exploratory
ability and causing a drastic slowdown;

3) They do not incorporate a temperature, hence they change the partition
function $Z=Z(\beta)$, and there is no guarantee that a quadratization
that recovers the original PUBO local field is possible. A general
fixed-temperature solver \cite{foos2025beyond} eventually samples
from a desired distribution, giving the relation $\braket{s}$ $\equiv\frac{1}{Z}\frac{\partial Z}{\partial h_{\mathrm{eff}}}$,
where $h_{\mathrm{eff}}$ is the effective field which always exists
in principle. This means that as long as the partition function is
available, and is preserved by a parameter transformation, an effective
field is retrievable from it. We will make use of this fact in sec.
\ref{sec:Effective-Interaction} to obtain the effective interaction
of the extra spins introduced and thus recover the original field.
For near-to-equilibrium solvers such as simulated annealers, or for
transformations that change $Z(\beta)$, exact equations of statistical
equilibrium are no longer attainable, and there is neither a guarantee
that an effective field exists nor that it can be used to recover
the original PUBO field. Furthermore, our quadratization properly
accounts for a general temperature $\beta$, and we leverage this
freedom to parallelize our algorithm at arbitrary temperature intervals,
to obtain a significant speedup.

Past works in quadratization introduce one additional spin for each
clause. This additional spin is connected to three original variables
\cite{10.1162/neco.1991.3.2.282}, which is leveraged to create a
partial alignment of the cost function or energy for one of the additional
spin values, say $+1$, yet not for the other. An ad-hoc penalty term
is then required to enforce the solver to pick the correct configuration.

In this work, we propose a quadratization that aims at preserving
the free energy (or partition function) of the original system, rather
than solely the energy. A statistically equivalent version of the
third-order term is created by introducing an ``extra'' spin $s_{e}$
with an external field $h_{e}$, and couple it to the three ``base''
(original) spins in the third-order interaction. Just as with the
classical quadratization scheme by Rosenberg \cite{rosenberg1972breves},
a mismatch in energy between configurations with equal base spins
exists in state space, however such a configuration will never be
sampled when equilibrium is attained by construction. Moreover, this
transform is exact in the sense that all the terms in the new Hamiltonian
are uniquely defined by the transform. This statistical equivalence,
known as position space renormalization group star-triangle transformation
\cite{PhysRevB.27.5678,Walker_2023}, has been known and used to great
effect in condensed matter physics as a calculation tool for thermodynamic
quantities on lattices \cite{PhysRevB.4.3174,KADANOFF1976359,10.1063/1.444839,RevModPhys.70.653,10.1007/3-540-08848-2_17}.
This transform preserves the partition function $Z=\sum_{s}e^{-\beta H(s)}$
by a partial summation over spins $\sum_{s_{e}}$, which guarantees
that statistical correlations of any order between spins are maintained.
Other works in the same spirit are ref.\cite{e20100739}, who utilized
gauge transforms on the spins rather than on parameters, however without
eliminating the third-order interaction; and ref. \cite{doi:10.1126/science.aab3326}
who were able to engineer every energy level, at a cost of a great
number of auxiliary spins.

We here show that we can evade the direct sampling of the extra spins
and run a Monte Carlo process for only the base spins that interact
in a pairwise manner, thus yielding a significant speedup in finding
the ground state: We demonstrate a generic property of the transform:
The extra spins are conditionally independent, that is, they are independent
of each other provided the base spins are fixed. This allows us to
trace them out exactly, thereby evading explicit sampling of their
states by the Markov chain. An even simpler, albeit worse performing,
version can be derived by taking the low temperature limit which leverages
a parallel update rule instead. The latter gives insight into the
energy spectrum that QUBO traverses, explaining its drastic slowdown.
This competitive gain in performance can be understood because in
the effective system the traced-out extra spins appear to be in an
exponentially large number of states simultaneously, and thus represent
a large search space whose sampling is no longer required.

We conclude by testing the performance of these algorithms with parallel
tempering \cite{Mueller2010,PhysRevLett.57.2607,doi:10.1143/JPSJ.65.1604},
an established method to accelerate fixed-temperature optimizers. 

\section{\label{sec:level1} Ising Mapping of the 3-SAT cost function}

A k-SAT problem, denoted here as $F$, of $N$ Boolean variables $x_{i}\in\{0,1\}_{i=1,\ldots,N}$
is constructed by forming the logical \textbf{AND} between $M$ different
clauses, where each clause $C_{l}$ is the logical \textbf{OR} between
$k$ different literals. In our study, the $k$ literals are chosen
randomly among the $N$ possible indices $i$, and for each $i$,
a literal $z_{i}$ is formed that equals $x_{i}$ or its negation
$\Bar{x}_{i}=1-x_{i}$ with probability one-half. We write this rule
as follows:
\begin{equation}
F=\bigwedge_{l=1}^{M}C_{l}=\bigwedge_{l=1}^{M}\quad\left(\bigvee_{n=1}^{\text{{k}}}z_{i_{n}}^{(l)}\right).
\end{equation}

If there exists an appropriate assignment of all values $x_{i}$ that
evaluates $F$ to TRUE, then $F$ is called satisfiable. Otherwise,
one searches for an assignment that satisfies the maximum amount of
clauses simultaneously, hence named a MAX-k-SAT optimization problem.
We will work with satisfiable problems consisting of clauses of length
$\text{{k}}=3$, that is, 3-SAT problems.

This optimization problem can be embedded in a statistical-mechanics
problem by interpreting the cost function, or number of violated clauses
$N_{VC}$, as the energy of a system obeying the Boltzmann distribution
with probability $p=e^{-\beta N_{VC}}/Z,$ with $Z$ being the partition
function. In the low-temperature limit $\beta=1/T\rightarrow\infty$,
the distribution becomes concentrated on the minima of $N_{VC}$,
and the original optimization setting is recovered. To ease notation
we make the change of variables $x_{i}\in\{0,1\}$ to $s_{i}=2x_{i}-1\in\{-1,1\}$
and the identification of $N_{VC}$ with the energy given by the Hamiltonian
$H^{P}$, so that the 3-SAT Hamiltonian reads \cite{PhysRevE.56.1357}
(see App.$~\ref{sec:Hamiltonian-of-the-3-SAT}$)

\begin{equation}
\begin{split}H^{P}(s)=\frac{M}{8}-\sum_{i=1}^{N}J_{i}s_{i}-\frac{1}{2}\sum_{i=1}^{N}\sum_{j=1}^{N}K_{ij}s_{i}s_{j}\\
-\frac{1}{6}\sum_{i=1}^{N}\sum_{j=1}^{N}\sum_{k=1}^{N}L_{ijk}s_{i}s_{j}s_{k}\,.
\end{split}
\label{eq:H_normal}
\end{equation}
We use the Monte Carlo (Glauber dynamics \cite{Glauber1963}) method
to sample from the Boltzmann distribution, where we select in each
time step one spin at random for update and update it according to
the rule

\begin{equation}
s_{i}\leftarrow\begin{cases}
+1 & \text{ with prob. }\mathscr{E}_{i}(s)=\left(1+e^{-2\beta h_{i}^{P}(s)}\right)^{-1}\\
-1 & \text{ with prob. }1-\mathscr{E}_{i}(s),
\end{cases}\label{eq:Glauber}
\end{equation}
where 
\begin{align}
h_{i}^{P} & =-\frac{1}{2}(H^{P}(s_{i}=+1)-H^{P}(s_{i}=-1))\label{eq:local_field}
\end{align}
is the local field on spin $s_{i}$. The sequence of states sampled
by this dynamics obeys the same statistics as in Glauber dynamics,
where individual spins are updated by a Poisson process; differences
only concern temporal correlations of spins. Problems with parameters
with ratio $\alpha\equiv M/N\approx4.28$ exhibit a phase transition
\cite{Mezard2002} and the longest time to achieve equilibrium, therefore
we consider this regime to test our algorithms.

\section{transformation to a Qubo problem}

\begin{figure*}
\includegraphics[scale=0.7]{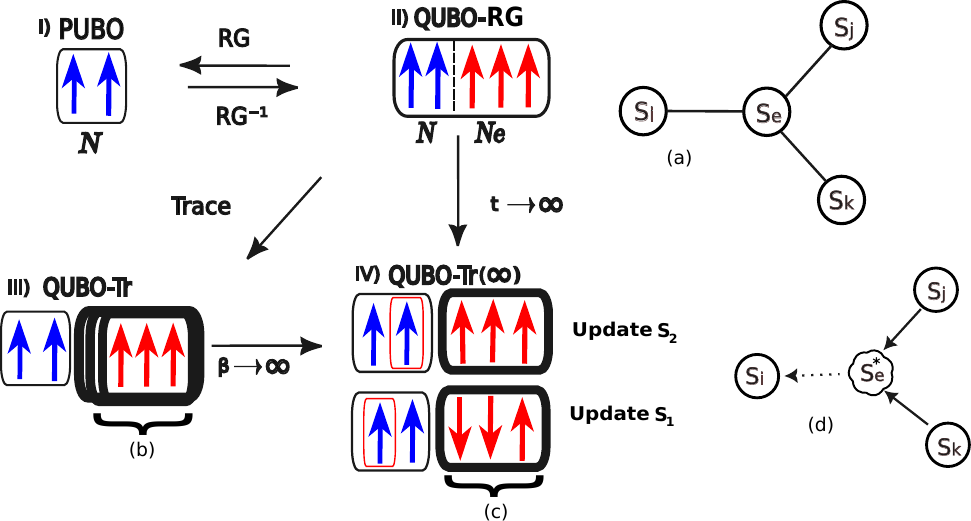}

\caption{\textbf{System Overview} \textbf{I)} Original system with third-order
interactions (PUBO) and $N$ base spins (blue). \textbf{II)} The inverse
RG transform, Eqs.$~\ref{eq:RG}$ and \ref{Hq}, maps PUBO to a system
with only pairwise interactions QUBO-RG, but with $N_{e}$ extra spins
(red) in addition.\textbf{ III)} Performing the trace over the extra
spins' configurations is done in QUBO-Tr, Eq.$~\ref{eq:dissect}$,
yielding the original PUBO local field, where the extra spins play
the role of an effective field acting on each base spin.\textbf{ IV)}
We link the $\beta\rightarrow\infty$ limit of the traced system with
the long runtime ($t\rightarrow\infty$) limit of QUBO-RG, which is
a novel way to interpret the concept of ``ruggedness'' in systems
with a fixed temperature, in Fig.$~\ref{fig:fraction_red}$. (a) Each
extra spin $s_{e}$ has a pairwise interaction with only three base
spins $s_{i}$, $s_{j}$, $s_{k}$, stemming from the original third
order interaction $L_{ijk}s_{i}s_{j}s_{k}$. (b) Extra spins are traced
out, which amounts to taking into account all of their $2^{N_{e}}$
configurations at once, in a superposition, and giving rise to an
additional effective field Eq.$~\ref{eq:dissect}$ acting on spin
$s_{i}$ to be updated. (c) In the limit of $\beta\to\infty$, the
superposition of extra spins becomes apparent: Depending on the base
spin to be updated (indicated by red rectangle, top vs bottom row
describe two different updates), the extra spins (red arrows) appear
to reside in a different state, described by their local field $\bar{h}_{e}\backslash_{i}$
in Eq.$~\ref{eq:parallel}$. (d) This local field $\bar{h}_{e}\backslash_{i}$
corresponds to a cavity field imposed on $s_{i}$, which is independent
of the value of $s_{i}$ and is interpreted as originating from a
fictitious extra spin $s_{e}^{*}$, see Eq.$~\ref{eq:fict}$.\label{fig:System-Overview}}
\end{figure*}

A technical challenge to physically implement the Glauber dynamics
of the PUBO system lies in the computation of the local field by \eqref{eq:local_field},
which requires the implementation of a third-order tensor $L_{ijk}$.
We therefore derive a representation of the problem here that only
employs pairwise interactions (Fig.$~\ref{fig:System-Overview}$)
and hence only requires matrices to be implemented physically. This
quadratization starts from the 3-SAT Hamiltonian in Eq.$~\ref{eq:H_normal}$,
$H^{P}[J,K,L]$, that we denote explicitly as PUBO from here on with
variables denoted by ``base spins'' $s_{b}$. We aim at a map to
a QUBO Hamiltonian $H^{Q}[J^{Q},K^{Q}]$ with additional ``extra
spins'' $s_{e}$ such that 
\begin{equation}
e^{-\beta H^{P}(\{s_{b}\})}=\sum_{\{s_{e}\}}e^{-\beta H^{Q}(\{s_{b},s_{e}\})}\,,\label{eq:RG}
\end{equation}
akin to an inverse renormalization step. In the following, the resulting
Hamiltonian $H^{Q}$ from this procedure will be referred to as QUBO-RG.
The construction of $H^{Q}$ follows two steps where at every step
we will reference the resulting effect of the partial summation in
Eq.$~\ref{eq:RG}$.

First, the star-triangle transformation \cite{PhysRevB.27.5678} is
used to generate two new parameters $h_{e}$ and $k$ that account
for one tensor element $L_{ijk}$ in Eq.$~\ref{eq:H_normal}$, by
the addition of one extra spin $s_{e}$. This step performs an inversion
of Eq.$~\ref{eq:P-term}$ in App.$~\ref{sec:Star-triangle-transform}$.
A partial summation of the degrees of freedom $s_{e}$ at this stage
spawns undesired first order $J'$ and second order $K'$ terms as
well, which yield an undesired partition function. This is shown with
the arrows from Eq.$~\ref{Hq}$ to Eq.$~\ref{eq:transform}$ below.
Therefore, a further correcting step is performed without spoiling
the structure that accounts for $L_{ijk}$.

For this, we observe that parameters multiplying spins \textit{not}
summed over remain invariant under a partial summation (see details
in \ref{sec:Invariant}). Therefore, in a second step, we add the
terms $(J-J')$ and $(K-K')$ to Eq.$~\ref{Hq}$ that pass through
unaffected by the partial summation $\sum_{s_{e}}$ down to Eq.$~\ref{eq:transform}$.
Hence, the undesired terms get canceled (`` = 0'') and the terms
that are left represent the rest of the 3-SAT Hamiltonian. The resulting
Hamiltonian $H^{Q}$ satisfies Eq.$~\ref{eq:RG}$ so that it has the
same partition function as $H^{P}$, but contains at most quadratic
terms. These interactions can be implemented by new matrices which
are element-wise corrections of the PUBO ones: $[J^{Q}]_{i}=[J-J']_{i}\text{ {and} }[K^{Q}]_{ij}=[K-K']_{ij}$.
The external fields $h_{e}$ and couplings $k$ of the extra spins
are encoded in the same matrices as the base spins, but placed in
a position of the matrices with larger line and column number, which
we denote as augmented $\overline{J}^{Q},\overline{K}^{Q}$ matrices.
We detail this encoding in App. $~\ref{sec:clauses}$.
\begin{widetext}
\begin{gather}
H^{Q}(s)={\normalcolor \textcolor{blue}{k}}(L)(s_{i}+s_{j}+s_{k})\tikzmark{s0}{\underbrace{s_{e}+\textcolor{blue}{h_{e}}(L)s_{e}}}\rlap{\ensuremath{\,}}+(J-\textcolor{red}{J'({\normalcolor L})})s_{i}+(K-\textcolor{red}{K'({\normalcolor L})})s_{i}s_{j}\label{Hq}\\[2em]
H^{P}(s)=\underbrace{(J'-\textcolor{red}{J'})}s_{i}+\underbrace{(K'-\textcolor{red}{K'})}s_{i}s_{j}+{Ls_{i}s_{j}s_{k}}+Js_{i}+Ks_{i}s_{j}\label{eq:transform}
\end{gather}

\begin{tikzpicture}[overlay,remember picture]	\draw[arrows=->]      ( $ (pic cs:s0) +(27.5pt,-2ex) $ ) --  
   ( $ (pic cs:s0) +(20pt,-7ex) $ );     
	\draw[arrows=->]     ( $ (pic cs:s0) +(27.5pt,-2ex) $ ) -- 
    ( $ (pic cs:s0) +(-36pt,-7.5ex) $ );     
	\draw[arrows=->]      ( $ (pic cs:s0) +(27.5pt,-2ex) $ ) --
     ( $ (pic cs:s0) +(84.5pt,-7.6ex) $ );     
	\node[anchor=north]     at ( $ (pic cs:s0) +(34pt,-12ex) $ )     
	{ = 0};
    \node[anchor=north]     at ( $ (pic cs:s0) +(-30pt,-12ex) $ )     
	{ = 0}; 
   \node[anchor=north]     at ( $ (pic cs:s0) +(-170pt,-2ex) $ )     
	{ $\sum_{ \{s_e\}}e^{-\beta H^Q} : $};

\end{tikzpicture}
\end{widetext}

We stress that this approach is completely general and independent
of the specific value of the parameters $J,K,L$ because a correspondence
can be set up for any value of $L$. Such a procedure is repeated
for every three-point interaction, which yields one extra spin $s_{e}$
each (see details in \ref{sec:clauses}).

\subsection{\label{sec:Ground}Ground-state Preservation}

As the expected value for a spin in equilibrium is $\braket{s_{i}}=\frac{{1}}{Z}\sum_{\{s_{b},s_{e}\}}s_{i}e^{-\beta H^{Q}}$,
the condition that the QUBO-RG Hamiltonian possesses the same partition
function under a renormalization, by Eq.$~\ref{eq:RG}$, is sufficient
to guarantee that the expectation values of base spins are preserved
in the transform giving Eq.$~\ref{eq:expet}$. Moreover, in the limit
$\beta\rightarrow\infty$, due to lack of thermal energy, only the
lowest-energy configurations $s_{\mathrm{PUBO}}^{GS}$ contribute
to the expectation, yielding Eq.$~\ref{eq:t0}$. This means that the
lowest-energy configurations are mapped one-to-one, as we show in
table$~\ref{tab:en_levels}$. In particular, it holds that

\global\long\def\myeq{\mathrel{\overset{\makebox[0pt]{\mbox{\normalfont{\tiny \sffamily RG}}}}{=}}}%
 
\global\long\def\myeqq{\mathrel{\overset{\makebox[15pt]{\mbox{\normalfont{\tiny \sffamily\ensuremath{\beta\to\infty}}}}}{=}}}%

\begin{gather}
\braket{s_{i}}_{\mathrm{PUBO}}\myeq\braket{s_{i}}_{\mathrm{QUBO}}\,,\label{eq:expet}\\
\braket{s_{i}}_{\mathrm{PUBO}}\myeqq s_{\mathrm{PUBO}}^{GS}\,.\label{eq:t0}
\end{gather}

\begin{table}[h!] \centering \renewcommand{\arraystretch}{1.3} \begin{tabular}{@{}lccc@{}} \toprule  & \textbf{PUBO} & \textbf{Rosenberg} & \textbf{QUBO-RG} \\ \midrule \textbf{Config.} & $-1\ -1\ -1$ & $\begin{array}{c} -1\ -1\ -1\ -1 \\[-0.3em]-1\ -1\ -1\ +1 \end{array}$ & $
\begin{array}{c} -1\ -1\ -1\ -1 \\[-0.3em] -1\ -1\ -1\ +1 \end{array}$ \\ \textbf{Energy} & $1$ & $
\begin{array}{c} 1 \\[-0.3em] 1 \end{array}$ & $
\begin{array}{c} 1.002 \\[-0.3em] 7.110 \end{array}$ \\ \textbf{Prob.} & $0.050$ & $
\begin{array}{c} 0.050 \\[-0.3em] 0.050 \end{array}$ & $\begin{array}{c} 0.050 \\[-0.3em] 0.000 \end{array}$ \\ \midrule \multicolumn{4}{c}{\large$\vdots$} \\ \midrule \textbf{Config.} & $1\ 1\ 1$ & $\begin{array}{c} 1\ 1\ 1\ -1 \\[-0.3em] 1\ 1\ 1\ +1 \end{array}$ & $\begin{array}{c} 1\ 1\ 1\ -1 \\[-0.3em] 1\ 1\ 1\ +1 \end{array}$ \\ \textbf{Energy} & $0$ & $\begin{array}{c} 0 \\[-0.3em] 3 \end{array}$ & $\begin{array}{c} 2.073 \\[-0.3em] 0.135 \end{array}$ \\ \textbf{Prob.} & $0.136$ & $\begin{array}{c} 0.136 \\[-0.3em] 0.007 \end{array}$ & $\begin{array}{c} 0.017 \\[-0.3em] 0.119 \end{array}$ \\ \midrule $\begin{array}{c} \langle s_1 \rangle \\[-0.3em] \langle s_2 \rangle \\[-0.3em] \langle s_3 \rangle \end{array}$ & $\begin{array}{c} 0.086 \\[-0.3em] 0.086 \\[-0.3em] 0.086 \end{array}$ & $\begin{array}{c} -0.005 \\[-0.3em] -0.005 \\[-0.3em] 0.153 \end{array}$ & $\begin{array}{c} 0.086 \\[-0.3em] 0.086 \\[-0.3em] 0.086 \end{array}$ \\ \bottomrule \end{tabular} \caption{Example of statistical quantities of interest in the Ising formulation with $\beta = 1$ of the problem $(x_1 \lor x_2 \lor x_3)$ for its PUBO, Rosenberg, and QUBO-RG approaches. For each possible configuration of spin variables $(s_1,s_2,s_3)$, the energy (E) and probability (Prob.) are grouped together as three rows. Different blocks correspond to different configurations. The energy values for Rosenberg come from ref.\cite{10558658}. The probability is calculated as $p= e^{-\beta E}/Z$, which is valid only in a statistical equilibrium condition. Final row summarizes the expectation value of each of the three involved spins, that is computed from information of all the blocks. Comprehensive table in Appendix \ref{tab:compare_rosen}.} \label{tab:en_levels} \end{table} 

From the last group of lines in table$~\ref{tab:en_levels}$, it is
clear that the Rosenberg transform preserves partially the energy
of a PUBO configuration, however does not preserve expectation values.
This is generally true for other methods that attempt maximal similarity
between energy levels, such as \cite{10937013}. In contrast, QUBO-RG
introduces configurations with different energy than any PUBO configuration,
yet because it preserves the free energy, it preserves expectation
values in a statistical equilibrium (see Eq.$~\ref{eq:free_en}$).
In terms of serving as a better optimization solver, we explore performance
implications and methods for further improvement in the following.

\section{\label{sec:Effective-Interaction}Effective Interaction of the Extra
Variables}

By construction, if one traces out the extra spins that is, one computes
the sum over them, as on the right hand side of Eq. \eqref{eq:RG},
one obtains the initial PUBO interaction. Moreover, we notice that
the local field on each extra spin $\bar{h}_{e}^{Q}$ depends solely
on the three base spins $\{s_{i},s_{j},s_{k}\}$ to which it is connected
(see \prettyref{fig:System-Overview} a)

\begin{equation}
\bar{h}_{e}^{Q}(s_{i},s_{j},s_{k})=k(s_{i}+s_{j}+s_{k})+h_{e}\,,\label{eq:local_extra_field}
\end{equation}
which means that the extra spins are mutually independent, provided
the base spins are fixed. This suggests that the extra spins' effect
can be traced out from the remaining interactions that render the
degrees of freedom mutually dependent, whilst performing the minimization.
This is the main idea that gives rise to the QUBO-Tr system (\prettyref{fig:System-Overview}
a). A direct calculation using Eqs.$~\ref{eq:local_field}$ and $\ref{eq:RG}$
yields for the local field $h_{i}^{P}$ that controls the update
step of base spin $s_{i}$ in the QUBO-Tr system:

\begin{equation}
h_{i}^{P}=h_{i}^{Q}(s_{b})+\frac{1}{2\beta}\sum_{\xi\in S_{e}(i)}\ln\frac{\cosh{\beta\Bar{h}_{e(\xi)}^{Q}(+1,s_{j},s_{k})}}{\cosh{\beta\Bar{h}_{e(\xi)}^{Q}(-1,s_{j},s_{k})}},\label{eq:dissect}
\end{equation}
where $h_{i}^{Q}(s_{b})=J_{i}^{Q}+({K^{Q}}\bold{s_{b}})_{i}$, with
$\bold{s_{b}}$ the array of $N$ base spins of the 3-SAT problem
and $S_{e}(i)$ is the set of indices of extra spins in the QUBO system
connected to spin $s_{i}$. For a clause to variable ratio $\alpha=4.28$
near the phase transition$S_{e}(i)$ has, on average, 13 elements
(that is, $3\alpha$, see \cite{PhysRevE.110.045308}). The effective
contribution of the extra spins, at every update of the Markov process,
amounts to a sum of nonlinear functions of the state of the $N$ base
spins. This sum arises from the conditional independence of the extra
spins. The term $h_{i}^{Q}(s_{b})$ contains the remaining (corrected)
pairwise interactions that are taken into account as in the QUBO system.
One can, therefore, bypass the extra spins and instead compute the
nonlinear term, to replicate exactly the PUBO field Eq.$~\ref{eq:local_field}$.
We highlight that the contribution from the extra spins is independent
of $s_{i}$ when calculating the field for updating $s_{i}$, and
that our nonlinear expression is the same as the forwarded field in
a cavity-method calculation \cite{Mezard2001} or in a message-passing
algorithm (e.g. belief propagation) as in \cite{9hx7-pzxw} and \cite{Braunstein2005}.
For such cavity methods to be exact, however, spins traditionally
need to span a tree configuration. The result above, in contrast,
is exact by construction in the current system (see Fig.$~\ref{fig:System-Overview}$
b).

For a potential hardware implementation it may be advantageous to
write the $\ln\cosh$ term from Eq.$~\ref{eq:dissect}$ using the
simpler $\operatorname{sign}$ function as the source of nonlinearity,
as detailed in App. $~\ref{subsec:Signum-approach}$.

\begin{figure}
\includegraphics[scale=0.4]{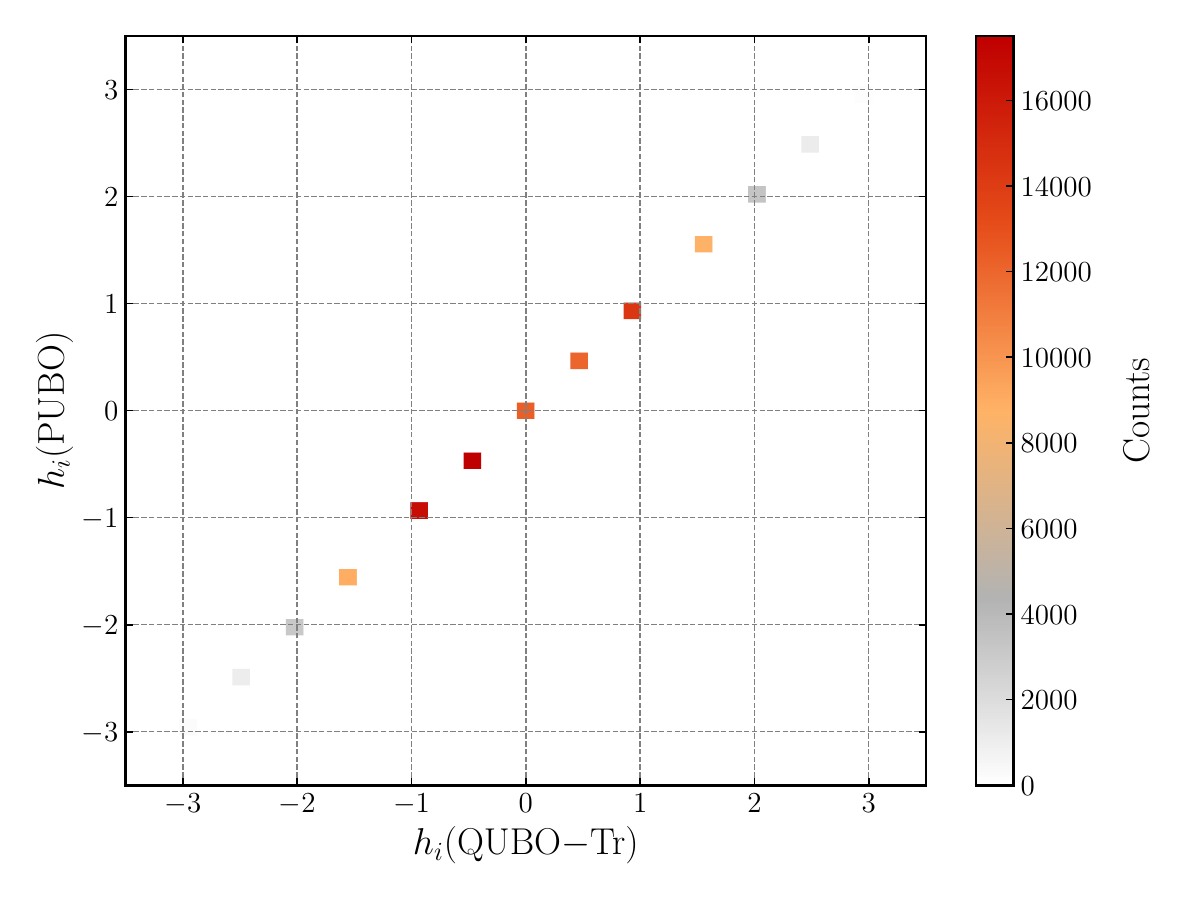}

\caption{Density plot of local fields of PUBO Eq.$~\ref{eq:H_normal}$ and
QUBO-Tr Eq.$~\ref{eq:dissect}$. Run on an $N=100$ instance for $\tau_{s}=10^{5}$
steps and temperature with $\beta=5$. At each step the two fields
are separately calculated for the same spin selected for update. See
App.$~\ref{sec:Comparison_QUBO_PUBO}$ for the comparison of QUBO-RG
and PUBO fields.}
\label{fig:density}
\end{figure}

In Fig.$~\ref{fig:density}$ we exemplify Eq.$~\ref{eq:dissect}$
and the range of values of the local field in PUBO. The two fields
coincide perfectly for arbitrary temperatures. In contrast to $h_{i}(\text{{PUBO}})$,
however, computing $h_{i}(\text{{QUBO-Tr}})$ does not require any
tensor operations or products of spins.

\subsection{Algorithm 1 - Bypass of extra spins: QUBO-Tr}

The method for calculating $h_{i}^{P}$ with Eq.$~\ref{eq:dissect}$
will require additional hardware resources, so we here develop an
efficient calculation by first expressing the field in Eq.$~\ref{eq:local_extra_field}$
by a vector-matrix multiplication where we order the extra spins:

\begin{equation}
\begin{pmatrix}\bar{h}_{e(1)}^{Q}\\
\bar{h}_{e(2)}^{Q}\\
\vdots\\
\bar{h}_{e(N_{Q}-N)}^{Q}
\end{pmatrix}=k\boldsymbol{\mathbb{\mathbb{G}}}\begin{pmatrix}s_{1}\\
s_{2}\\
\vdots\\
s_{N}
\end{pmatrix}+\begin{pmatrix}h_{e(1)}\\
h_{e(2)}\\
\vdots\\
h_{e(N_{Q}-N)}
\end{pmatrix}\,,\label{eq:M}
\end{equation}
where matrix $\boldsymbol{\mathbb{\mathbb{G}}}$ has elements given
by (the indicator $1_{(\bar{h}_{e(i)}^Q \text{ depends on } s_j)}$
function) $\bold{\mathbb{{G}}}_{ij}=1$ if $\bar{h}_{e(i)}^{Q}$ depends
on $s_{j}$ and $0$ otherwise, with dimensions $(N_{Q}-N)\times N$
(see App.$~\ref{sec:clauses}$). In a given solver, this matrix would
then be computed once at the start of the solution process. It encodes
the indices of the extra spins involved in the update of spin $s_{i}$
in column $\boldsymbol{\mathbb{\mathbb{G}}}[\colon,i]$, which contains
about 13 (or $3\alpha$) non-zeros on average. We compute on each
line in the column matrix of Eq. $\ref{eq:M}$ the local field on
each extra spin $\bar{h}_{e(\xi)}^{Q}$. For each local field $h_{i}^{P}$
of Eq.$~\ref{eq:dissect}$, the necessary terms $\bar{h}_{e(\xi)}^{Q}(\pm1,s_{j},s_{k})=k(s_{j}+s_{k})+h_{e}\pm k$
(Eq.$~\ref{eq:local_extra_field}$) are thus also computed using both
the local fields on extra spins and the relevant base spin: $\bar{h}_{e(\xi)}^{Q}(\pm1,s_{j},s_{k})=\bar{h}_{e(\xi)}^{Q}-ks_{i}\pm k$;
see line 16 in the pseudocode of Appendix $\ref{subsec:Look-up-table-approach}$.
The latter term can only have six different values in total (three
for $(s_{j}+s_{k})$ and two for the sign of $h_{e}$), so one possibility
is to fetch them from two look-up tables of size 2 $\times$ 3 each
($\mathbb{LUP}$ and $\mathbb{LUM}$), which are again constructed
once at the start -- see App.$~\ref{subsec:Look-up-table-approach}$.
In the following pseudocode however, we show a simpler version which
allows for easier counting of operations. We stress that, in addition
to the usual pairwise interactions, the number of elements to be summed
over is an intensive quantity and depends solely on the fixed ratio
$\alpha=M/N$.

\begin{algorithm}[H]  \begin{algorithmic}[1] \Statex \textbf{Construction of matrix} $\boldsymbol{\mathbb{G}}$: \For{$i \in \{0,\dots,N_Q - N - 1\}$}     \For{$j \in \{0,\dots,N - 1\}$}         \State $\boldsymbol{\mathbb{G}}[i,j] = 1_{(\bar{h}_{e(i)}^Q \text{ depends on } s_j)}$     \EndFor \EndFor \Statex  \Statex \hrulefill \Statex \textbf{Monte Carlo} \State \textbf{Input:} $\boldsymbol{s}$, $J^Q$, $K^Q$, $\boldsymbol{\mathbb{G}}$, $T_{\text{sim}}$, array of external fields $h_e$, $k$  \For{$t \leq T_{\text{sim}}$}     \State Pick one spin $s_i$ at random among $N$ possible ones.     \State $h_i = J^Q[i] + K^Q[i,:] \cdot \boldsymbol{s}$      \State $h_e^Q = \boldsymbol{\mathbb{G}}\cdot \boldsymbol{s} + h_e$      \State $h_e^{Q+} = h_e^{Q} + \boldsymbol{\mathbb{G}}[:, i](k-ks_i)$     \State $h_e^{Q-} = h_e^{Q} + \boldsymbol{\mathbb{G}}[:, i](-k-ks_i)$     \State $h_i = h_i + \boldsymbol{\mathbb{G}}[:, i]^T\cdot\frac{1}{2\beta}\left(\ln{\cosh h_e^{Q+}} - \ln{\cosh h_e^{Q-} } \right)$           \State Update selected spin $s_i$      \State Increment $t$ \EndFor \end{algorithmic} \caption{QUBO-Tr: Modified Glauber dynamics with correction to local fields} \label{alg:1} \end{algorithm}

In a potential implementation, this amounts to a better memory usage
compared to storing a 3-tensor $L_{ijk}$ of size $N^{3}$, as each
of the objects used $J^{Q},K^{Q},\ensuremath{\boldsymbol{\mathbb{\mathbb{G}}}}$
scales at most with $N^{2}$ in size and is sparse for 3-SAT problems.
Moreover, implementing the vector operations for each spin update
can be performed efficiently in a hardware solver, for example by
the use of memristive crossbar arrays \cite{article_cai}.

\subsection{Solver Performance}

\begin{figure}[H]
\includegraphics[width=1\columnwidth]{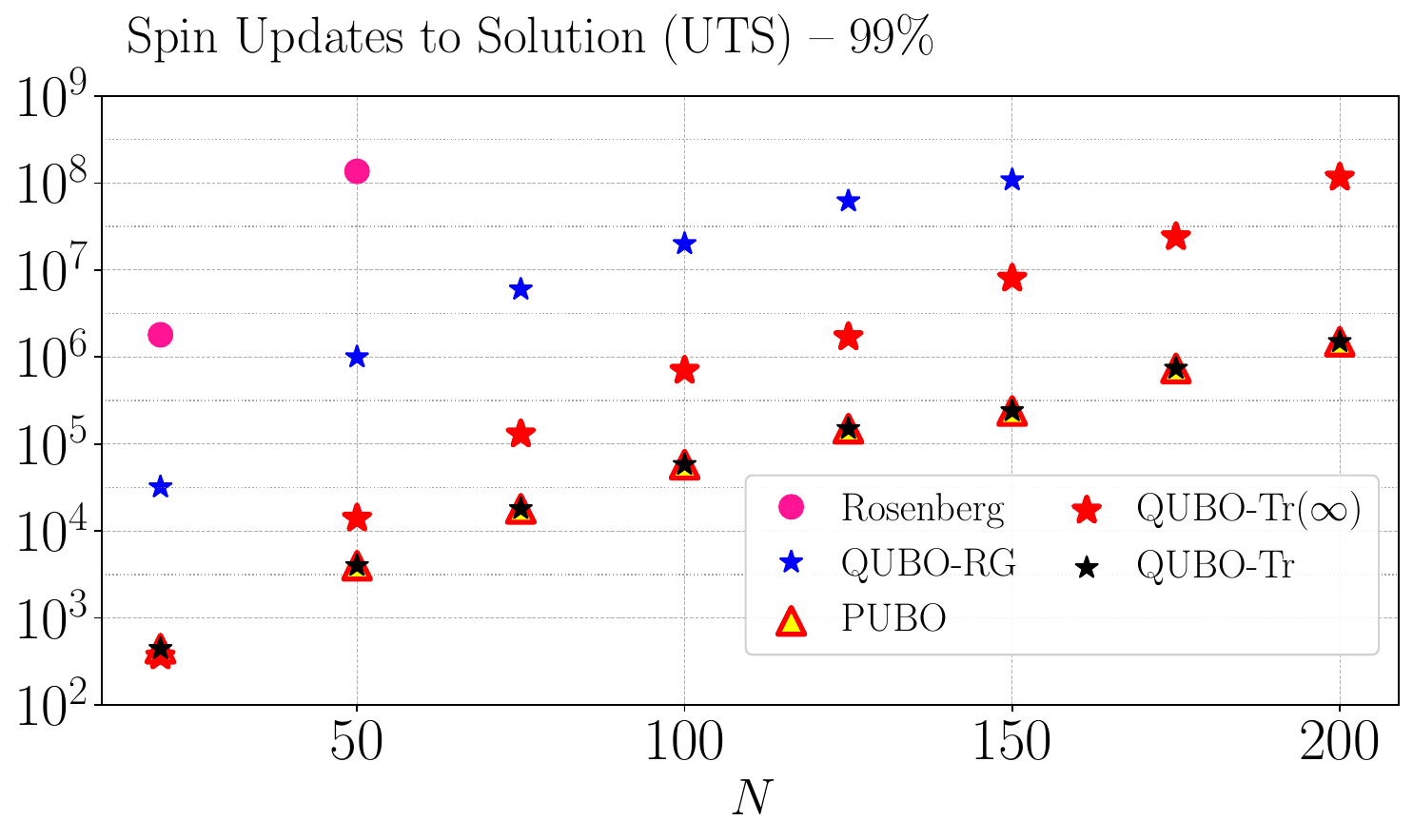}

\caption{Performance comparison between the classical quadratization technique
by Rosenberg \cite{10558658} and PUBO Eq.$~\ref{eq:H_normal}$, and
our novel approaches: QUBO-RG from Eq.$~\ref{Hq}$, QUBO-Tr($\infty$)
from Eq.$~\ref{eq:parallel}$, and QUBO-Tr from Eq.$~\ref{eq:dissect}$;
We show values up to $10^{8}$ due to corresponding runtimes being
exceedingly large. The parallel tempering heuristic \cite{PhysRevLett.57.2607}
is used to accelerate the spin updates needed to achieve a solution.
It consists of running our fixed-temperature solvers at many temperatures
in parallel (each temperature is a replica) and properly exchanging
temperatures after each run. Spin flips calculated in parallel were
not counted in the UTS. For each solver, 12 replicas are used, hence
there are 12 update steps performed in parallel, one for each parallel
tempering replica. The temperature schedule is $\beta=3n,\quad\text{{with}}n\in\{1,2,...,12\}$,
and a single swap of temperatures is performed. The two runs at a
given temperature, one before and one after the swap, were included
in UTS. See App.$~\ref{sec:pt_schedule}$ for a schedule justification
and details on the heuristic. Problem instances are random CSPs, retrieved
from the data set SATLIB ref.\cite{hoos_satlib} number \textit{uf$N$-[901,1000]}.
$\mathrm{UTS}_{99}=\tau_{s}\log(0.01)/\log(1-p_{s})$ determined from
the success probability $p_{s}$ which is estimated by the ratio of
successful solutions among 100 attempts, each started from a random
initial condition.\label{fig:compare_mo}}
\end{figure}

In Fig.$~\ref{fig:compare_mo}$ we plot an established measure of
the number of Monte Carlo steps (spin updates) needed to obtain a
solution with $99\%$ certainty: $\mathrm{UTS}_{99}=\tau_{s}\frac{\log(0.01)}{\log(1-p_{s})}$
\cite{Pedretti2025}. Here $\tau_{s}$ is the total number of Monte
Carlo steps each corresponding to one spin update. The probability
of success $p_{s}$ is estimated by running the algorithm 100 times
with random initial conditions on a fixed problem instance, and determining
the ratio of successful runs that reached the ground state. This is
repeated for 100 different instances, yielding a value of $\mathrm{UTS}_{99}$
for each problem instance. The value plotted is the median of all
values of $\mathrm{UTS}_{99}$ obtained. In Fig.$~\ref{fig:compare_mo}$
we observe QUBO-Tr has the same performance as PUBO, as expected due
to our exact local field construction \ref{eq:dissect}. QUBO-RG,
in blue, requires even more iterations, but still has a much better
performance compared to the Rosenberg algorithm \cite{rosenberg1972breves},
which is the classical quadratization technique.

From our UTS values we find the same pattern as those of ref. \cite{10558658}.
Namely their worst and best highly engineered, parallel, solvers are
comparable in order of magnitude to our worse (QUBO-RG) and best (QUBO-Tr)
points respectively. There are however stark differences, as the former
\cite{10558658} uses a Simulated-Annealing solver.

\subsection{\label{subsec:Algorithm-2--}Algorithm 2 - Parallel update scheme
in the low temperature limit: QUBO-Tr ($\infty$)}

If in hardware implementations one wishes to avoid the calculation
of the nonlinear term of Eq.$~\ref{eq:dissect}$ and one does not
want to use lookup tables, one may use an approximation instead. In
the $\beta\rightarrow\infty$ limit the nonlinear terms in Eq.$~\ref{eq:dissect}$
converge to
\begin{align*}
\frac{1}{2\beta}\,\ln\frac{\cosh{\beta\Bar{h}_{e(\xi)}^{Q}(+1,s_{j},s_{k})}}{\cosh{\beta\Bar{h}_{e(\xi)}^{Q}(-1,s_{j},s_{k})}} & \stackrel{\beta\to\infty}{\to}k\,s_{e}^{\ast}\,,
\end{align*}
where $s_{e}^{\ast}\in\{-1,1\}$. Here $s_{e}^{\ast}$ plays the role
of a fictitious spin $s_{e}^{*}$ giving a similar form to a QUBO-RG
local field (see Eq.$~\ref{Hq}$ and Figs.$~\ref{fig:System-Overview}$c,d):

\begin{equation}
h_{i}^{P}\overset{\beta\rightarrow\infty}{=}h_{i}^{Q}(s_{b})+k\sum_{\xi\in S_{e}(i)}s_{e(\xi)}^{*}\,,\label{eq:fict}
\end{equation}
with the crucial difference that, at any given step, all the $s_{e}^{*}$
are updated deterministically, and simultaneously, according to a
local field $\bar{h}_{e}\backslash_{i}$:

\begin{equation}
\begin{cases}
s_{e}^{*}=\text{{sgn}}(\bar{h}_{e}\backslash_{i})\\
\bar{h}_{e}\backslash_{i}=k(s_{j}+s_{k})+h_{e}.
\end{cases}\label{eq:parallel}
\end{equation}

\begin{figure*}
\includegraphics[scale=0.5]{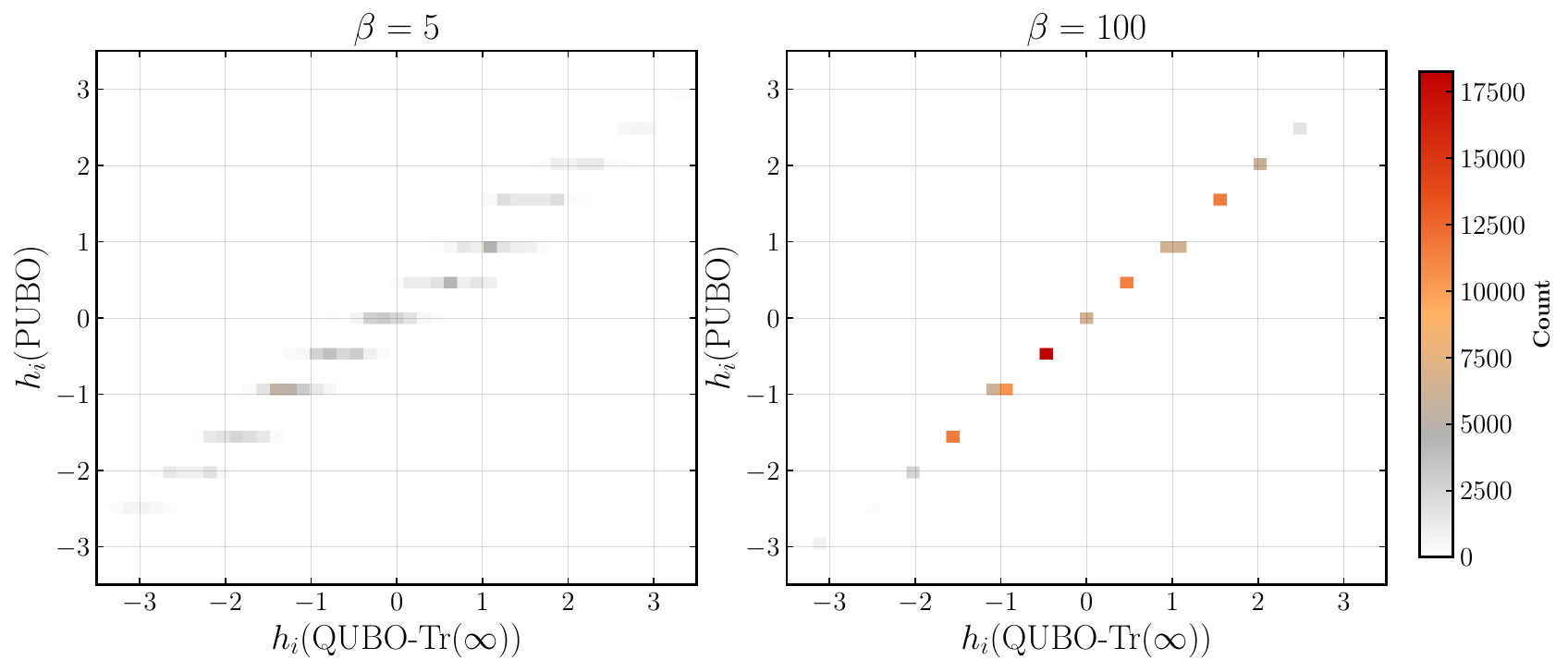}

\caption{Density plot of local fields of PUBO Eq.$~\ref{eq:H_normal}$ and
QUBO-Tr($\infty$) Eq.$~\ref{eq:fict}$. Run on an $N=100$ instance
for $\tau_{s}=10^{5}$ steps, where at each step the two fields are
separately calculated at the spin selected for update, with $\beta=5$
(left) and $\beta=100$ (right), respectively.}
\label{fig:quenched}
\end{figure*}

In Fig.$~\ref{fig:quenched}$ we plot how closely related the PUBO
field is to this approximated field, as a function of temperature.

The performance of QUBO-Tr($\infty$), which takes the limit $\beta\rightarrow\infty$,
is evaluated in Fig.$~\ref{fig:compare_mo}$ in red. It avoids the
nonlinear function $\ln\cosh$, however more iterations are needed
to reach a solution, relative to the effective interaction.

\section{Dynamics of the extra spins due to conditional independence}

\begin{figure}[H]
\includegraphics[width=1\columnwidth]{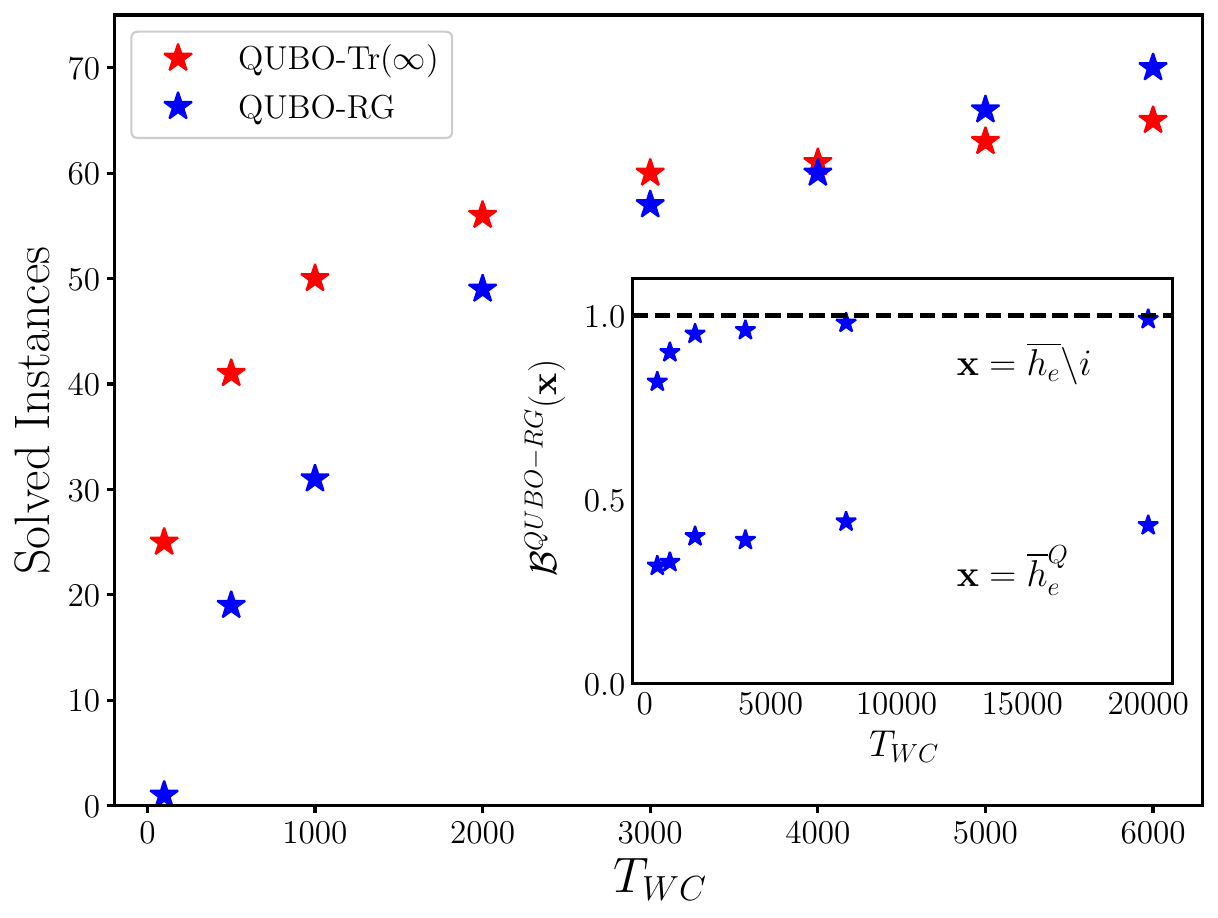}\caption{Comparison of QUBO-RG and QUBO-Tr($\infty$). Both algorithms are
applied to $N=100$ base spin problems for a range of temperatures
in parallel, averaged over 100 instances and 100 random initial conditions
each. The number of spin updates for each solver $T_{sim}^{\mathrm{QUBO-RG}}$and
$T_{sim}^{\mathrm{QUBO-Tr}(\infty)}$ is calculated as $T_{sim}^{\mathrm{QUBO-RG}}=T_{WC}\times N_{\mathrm{QUBO-RG}}$
and $T_{sim}^{\mathrm{QUBO-Tr}(\infty)}=T_{WC}\times N_{\mathrm{\mathrm{QUBO-Tr}(\infty)}},$
where $T_{WC}$ is the wall-clock steps and $N_{QUBO-RG},N_{QUBO-Tr(\infty)}$
are the number of spins that each solver runs on respectively. Differences
for long runtime between the curves are due to different local fields
prior to the final time step. The inset explains the similarity of
the curves at the plotted scale, where a QUBO-RG run is applied at
one problem with $N=100$ and $\beta=10$. During such a run, we keep
track of the quantity $\mathcal{B}^{\mathrm{QUBO-RG}}(\bar{h}_{e}\backslash_{i})$,
which counts the fraction of update steps of spin $i$ during which
all its connected extra spins have a configuration predicted by $\mathrm{sgn}(\bar{h}_{e}\backslash_{i})$;
such extra spins thus follow the $\beta\to\infty$ dynamics \eqref{eq:parallel}.
The fraction $\mathcal{B}^{\mathrm{QUBO-RG}}(\overline{h_{e}}^{Q})$
counts the fraction of update steps of spin $i$ during which all
its connected extra spins have a configuration predicted by $\mathrm{sgn}(\bar{h}_{e}^{Q})$.
As an example, if 15 out of the 16 extra spins connected to a base
spin are correctly aligned, this does not increase the count, see
App.$~\ref{sec:qubo_dynamics}$.\label{fig:fraction_red}}
\end{figure}

There exists a link between QUBO-Tr ($\infty$) and QUBO-RG, which
we detail below and show numerically in Fig.$~\ref{fig:fraction_red}$
(and Fig.$~\ref{fig:System-Overview}$ $(t\rightarrow\infty)$). As
the approaches use different numbers of spins, the time to solution
on a conventional computer with a single thread is expected to scale
in proportion to this number, since each spin needs to be selected
for update in turn. If implemented on a physical hardware, updates
of spins can be performed in parallel. As we envision such hardware
implementations as a future target, we here normalize the run time
by the total number of spins, which we denote as \textquotedblleft wall
clock steps\textquotedblright . This normalization, in addition, allows
us to compare the properties of the corresponding energy landscapes,
in particular their ruggedness.

While in Algorithm 2 one immediately sets the $s_{e}^{*}$ to the
value $\mathrm{sgn}(\bar{h}_{e}\backslash_{i})$, in the QUBO-RG case
the convergence of the extra spins takes about $10^{4}$ wall-clock
steps. This is reflected in the increased run time of QUBO-RG compared
to $\text{{\ensuremath{\mathrm{QUBO-Tr}}(\ensuremath{\infty})}}$
in order to reach the same number of solved instances, as shown in
Fig.$~\ref{fig:fraction_red}$. For long runtime, an equilibrium settles
where the extra spins point in a direction independent of $s_{i}$
(the direction given by $\bar{h}_{e}\backslash_{i}$), where $s_{i}$
is the base spin selected for update. This is a feature of conditional
independence: if the base spins do not influence the orientation of
the extra spins, the extra spins in turn cannot communicate with each
other, as they have interactions only with base spins.

We now argue that the dynamics (or landscape) of the base spins in
QUBO-RG eventually resembles the $\beta=\infty$ dynamics of QUBO-Tr.
Eqs.$~\ref{eq:fict}$ and \ref{eq:dissect} share the same first term
$h_{i}^{Q}$, up to a small differences due to parameters that vanishes
for high $\beta$, which is also the base-spin contribution in a QUBO-RG
update (Eq.$~\ref{Hq}$). Therefore, the degree to which the local
field of QUBO-RG differs from the local field of QUBO-Tr($\infty$)
is measured by $\mathcal{B}^{\mathrm{QUBO-RG}}(\bar{h}_{e}\backslash_{i})$,
which counts the fraction of QUBO-RG base spins with the same field
as in QUBO-Tr($\infty$). As this fraction eventually approaches unity
(Fig.$~\ref{fig:fraction_red}$), QUBO-RG eventually has equal local
fields as QUBO-Tr($\infty$). In sum, QUBO-RG eventually updates its
spins based on these same ($\beta=\infty$) local fields despite operating
at non-zero temperature. In App.$~\ref{sec:qubo_dynamics}$, we show
another version of this result. In this sense, the landscape of QUBO-RG
and its imposing slowdown is explained by its correspondence to the
$\beta\rightarrow\infty$ limit of the PUBO landscape.

A similar strategy of engineering a configuration of extra spins $s_{e}$
(or $s_{e}^{*}$) in order to replicate a local field $h_{i}^{P}$,
which we do in a limit of Eq.$~\ref{eq:dissect}$, was pursued for
another quadratization in ref. \cite{10937013}. The latter includes
a large amount of search steps for every Monte Carlo step, though.
Our Algorithm 2 is physically grounded and gives a direct prescription
as a function of the state, evading the need for a search in a vast
configuration space.

\section{\label{sec:Solver-Performance}Cost of Computation}

In order to account for the cost of computation of our algorithms,
we estimate the total number of operations performed in each, yielding
the computations-to-solution $\mathrm{CTS}_{99}$ with the relation
(\ref{eq:ots}):

\begin{equation}
\mathrm{CTS}_{99}=\mathrm{UTS}_{99}\times\frac{\text{{Computations}}}{\text{{Spin Update}}}\times\text{{Replicas}}\,.\label{eq:ots}
\end{equation}

With regard to third-order contributions, the original PUBO solver
must compute products $L_{ijk}s_{j}s_{k}$, while our approach computes
an equal number of either $\ln\cosh(s_{j}+s_{k}+h_{e}\pm k)$ or sign($s_{j}+s_{k}+h_{e}$)
terms, evading the utilization of the tensor $L_{ijk}$ and products
of spins. We summarize this extra factor in table$~\ref{tab:solver_comparison}$
(for details see App.$~\ref{subsec:Diagonal-matrix-approach}$). Although
the actual performance depends on the underlying hardware, this shows
that no exponential overhead is hidden in the correction term of Eq.$~\ref{eq:dissect}$.

\begin{table}[h!] 
\centering 
\begin{tabular}{@{}l c c@{}}
\toprule 
\textbf{Solver} & \textbf{Compt./Spin Update} & \textbf{Replicas} \\ 
\midrule 
PUBO                 & $21\alpha$ & 12 \\ 
QUBO-Tr              & $72\alpha$ & 12 \\ 
QUBO-Tr($\infty$)    & $72\alpha$ & 12 \\ 
QUBO-RG / Rosenberg  & $18\alpha$ & 12 \\ 
\bottomrule 
\end{tabular} 
\caption{Comparison of solver complexities and third-order computations. The values assume only non-zero elements are counted. The total number of Operations per iteration is counted for every cycle from App.~\ref{subsec:Diagonal-matrix-approach} and multiplied by the total number of replicas.}
\label{tab:solver_comparison} 
\end{table}

\begin{figure}[H]
\includegraphics[width=1\columnwidth]{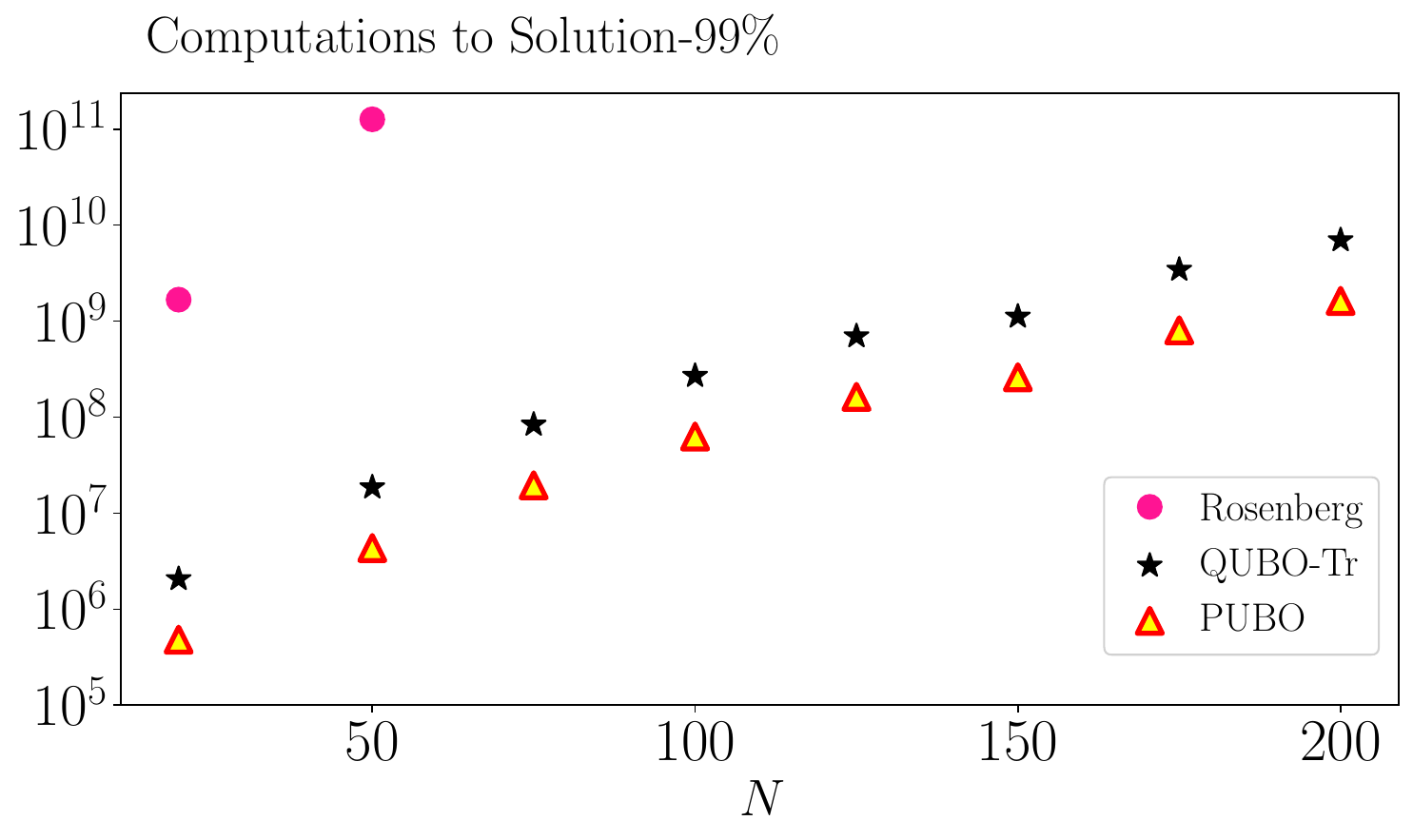}\caption{Number of computational operations to solution. The total number is
accounted by multiplying UTS-99 from Fig.$~\ref{fig:compare_mo}~$
with the computations per spin update derived in Table$~\ref{tab:solver_comparison}$.
Results show a constant overhead (problem size independent) for QUBO-Tr
compared to native PUBO, but importantly the same scaling behavior,
unlike the classical Rosenberg scheme.}
\label{fig:tts}
\end{figure}

In Fig.$~\ref{fig:tts}$ we plot the UTS data from Fig.$~\ref{fig:compare_mo}$
now multiplied by the total amount of operations. Using the Algorithm
\ref{alg:1}, QUBO-Tr does not superimpose with PUBO, although both
solvers still exhibit the same scaling, indicating that QUBO-Tr may
offer hardware advantages without a significant loss in performance.

\section{Conclusion}

We here develop a real-space renormalization approach that allows
us to reduce the order of interaction in Ising Hamiltonians, while
preserving the free energy. The technique is generic and by conserving
the free energy also conserves all statistical properties of the degrees
of freedom. The method comes at the expense of introducing additional
Ising spins. We are, however, able to show that this effective low-order
Hamiltonian possesses the useful property of conditional independence
of the newly introduced spins. This property allows us to again reduce
the degrees of freedom to the original ones. We show that the remaining
degrees of freedom undergo the identical Markov dynamics as the original
problem, but allows for more efficient implementations that circumvent
the need for costly higher order tensor operations.

We exemplify this generic technique on 3-SAT, an NP-complete combinatorial
optimization problem with polynomial interactions (PUBO), for which
we obtain novel solver techniques. Here the approach effectively reduces
the order of the interaction from third (PUBO) to second order (QUBO).
While previous attempts have been plagued by a drastic slow down in
the minimization, we here circumvent this problem by employing the
renormalization group concept. We further accelerate the minimization
of the different energies considered via the parallel tempering heuristic,
amenable to minimizers of fixed temperature.

In contrast to previous attempts which focused solely on the mapping
of the energy, this approach allows us to take into account the temperature-dependent
entropy in addition to the energy, hence the free energy. We show
how to obtain an equivalence between QUBO and PUBO by introducing
extra spins. Minimizing this (QUBO-RG) system already yields improved
performance relative to the classical Rosenberg algorithm of reducing
the order \cite{rosenberg1972breves}, however, sampling of extra
spins still takes place.

Crucially, a generic property of the presented approach is that the
additional spins are conditionally independent when the base spins
are fixed, which allows us to trace out these extra spins to obtain
an effective Monte Carlo process for the original spins alone (QUBO-Tr).
This not only evades the need to update the extra spins, but also
completely replicates the local fields of the initial 3-SAT problem,
hence giving a guarantee of equal Monte Carlo steps to reach the ground
state. This algorithm requires a nonlinear computation, for instance
a sign operation. The computation cost is estimated to amount to a
constant overhead with respect to PUBO, in stark contrast to the Rosenberg
algorithm \cite{rosenberg1972breves}. A key benefit of the presented
approach is that many modern Ising machines, such as Fujitsu DA \cite{article_fj},
Coherent Ising Machines \cite{doi:10.7566/JPSJ.92.044002,doi:10.1126/sciadv.abh0952},
Toshiba's second-order solver \cite{doi:10.1126/sciadv.abe7953},
Oscillator-based \cite{English2022}, quantum-inspired \cite{Jiang2023}
and other analog Ising machines \cite{Kalinin2018,deprins2025incorporatehigherorderinteractionsanalog}
are physically incapable of higher-order interactions. Thus, QUBO-Tr
offers a hardware realizable mapping to take advantage of the physics-based
solvers.

In addition to computational advantages, the physical insights into
QUBO-Tr provide another perspective into the concept of ruggedness
of the energy landscape. After a sufficient number of Markov steps,
the energy spectrum derived in the $\beta\rightarrow\infty$ limit
of the QUBO-Tr approach predicts increasingly well the distribution
of extra spins in QUBO-RG. This suggests that the energy landscapes
of these two systems show a comparable degree of ruggedness. Even
though QUBO-RG operates at non-zero temperature, it leads to the same
local fields that appear in the PUBO problem for long runtimes in
the $\beta\rightarrow\infty$ limit. This phenomenon could be explained
by a large gap between lower and excited states in the QUBO-RG spectrum,
which prohibits the state to visit higher-energy levels, despite thermal
fluctuations.

Moreover, the method of order reduction is applicable to general third-order
Ising Hamiltonians. It is thus applicable to other NP-complete problems
that can be expressed as a third-order Ising model. Among such additional
problems that can now be quadratized efficiently are the Parity constraint
satisfaction problem \cite{DriebSchon2023parityquantum}, 3-XOR-SAT
\cite{PhysRevLett.130.220601}, exact cover by 3-sets \cite{CHANG2003263},
3-Dimensional Matching \cite{article}, set cover with pairs \cite{Cao2016},
graph coloring \cite{Singh2025}, Hamiltonian cycles \cite{PhysRevX.14.031005}
and integer linear programming \cite{Karimzadehgan2012}.

We uncover a link between Belief Propagation and Markov-Chain Monte
Carlo via the inverse renormalization group. The former is a message
passing algorithm which exhibits difficulty when the graph of the
system contains loops. Such a difficulty is lifted in our approach,
as the Belief propagation step here becomes exact and amounts to a
Markov step. This property is independent of the loops in the problem
graph.

The methods presented here are likewise applicable to reduce the order
of 4-SAT problems, with the technical difference that in such a context
no inclusion of the external field $h_{e}$ is necessary since the
quartic interaction $Q_{ijkl}s_{i}s_{j}s_{k}s_{l}$ does not break
the up-down symmetry, as we found in 3-SAT. The applications of the
RG inversion to even higher-order contexts is left for future research.

In summary, we demonstrate here that theoretical physics techniques
may be employed to engineer the energy spectra of very general systems,
allowing for increased flexibility in physical implementations of
solvers with high efficiency, and contribute to our physical understanding
of the nature of combinatorial optimization problems.
\begin{acknowledgments}
A. C. is grateful to Mohammad Hizzani (PGI-14) for insightful suggestions
and technical help. A.C. received intramural funding as Vernetzungsdoktorand:
Physics and network-based combinatorial optimization. Open access
publication funded by the Deutsche Forschungsgemeinschaft (DFG, German
Research Foundation) -- 491111487. The authors gratefully acknowledge
the computing time granted by the JARA Vergabegremium and provided
on the JARA Partition part of the supercomputer JURECA at Forschungszentrum
J\"ulich (computation grant JINB33), and J\"ulich Supercomputing Centre.
(2018). JURECA: Modular supercomputer at J\"ulich Supercomputing Centre.
Journal of large-scale research facilities, 4, A132. http://dx.doi.org/10.17815/jlsrf-4-121-1
\cite{Julich2021}.

\providecommand{\noopsort}[1]{}\providecommand{\singleletter}[1]{#1}%

\end{acknowledgments}

\onecolumngrid

\section{Appendices}

\subsection{Hamiltonian of the 3-SAT problem\label{sec:Hamiltonian-of-the-3-SAT}}

The tensors $J,K,L$ in Eq. \eqref{eq:H_normal} are explicitly given
as

\begin{equation}
\left\{ \begin{aligned} & J_{i}=\frac{1}{8}\sum_{l=1}^{M}\Delta(l,i)\\
 & K_{ij}=-\frac{1}{8}(1-\delta_{ij})\sum_{l=1}^{M}\Delta(l,i)\Delta(l,j)\\
 & L_{ijk}=\frac{1}{8}(1-\delta_{ijk})\sum_{l=1}^{M}\Delta(l,i)\Delta(l,j)\Delta(l,k)
\end{aligned}
\right.,\label{eq:tensors}
\end{equation}
where

\[
\Delta(i,j)=\left\{ \begin{array}{cc}
1 & \text{if \ensuremath{x_{i}} appears in \ensuremath{\ensuremath{C_{j}}}}\\
-1 & \text{if \ensuremath{\overline{x}_{i}} appears in \ensuremath{C_{j}}}\\
0 & \text{if neither \ensuremath{x_{i}} nor \ensuremath{\overline{x}_{i}} appear in \ensuremath{C_{j}}}.
\end{array}\right.
\]

\subsection{Star-triangle transform\label{sec:Star-triangle-transform}}

Explicitly, the transform from Eq. \eqref{Hq} to Eq. \eqref{eq:transform}
reads

\begin{equation}
\left\{ \begin{aligned}-\beta H & =k(s_{1}+s_{2}+s_{3})s_{4}+hs_{4}\\
-\beta H' & =3K_{0}^{\prime}+h^{\prime}(s_{1}+s_{2}+s_{3})+K^{\prime}(s_{1}s_{2}+s_{1}s_{3}+s_{2}s_{3})+Ls_{1}s_{2}s_{3}
\end{aligned}
\right.\label{eq:stt}
\end{equation}
with $e^{-\beta H'}=\sum_{s_{4}}e^{-\beta H}$. There appears a third
order term after a partial summation over spin $s_{4}$ because the
field $h$ breaks the up-down symmetry, and all the terms consistent
with this property must be included in the resulting system. We note
that there corresponds one extra spin $s_{4}$ (or $s_{e}$) per unique
$L$ term. The explicit relations are:

\begin{equation}
8\beta L=\ln\left(\frac{\cosh\beta(3k+h)}{\cosh\beta(3k-h)}\right)+\ln\left(\frac{\cosh\beta(k-h)}{\cosh\beta(k+h)}\right)^{3}\label{eq:P-term}
\end{equation}

\begin{equation}
8\beta K^{\prime}=\ln\left(\frac{\cosh\beta(3k+h)}{\cosh\beta(k+h)}\frac{\cosh\beta(3k-h)}{\cosh\beta(k-h)}\right)\label{eq:kp}
\end{equation}

\begin{equation}
8\beta h^{\prime}=\ln\left(\frac{\cosh\beta(3k+h)}{\cosh\beta(3k-h)}\frac{\cosh\beta(k+h)}{\cosh\beta(k-h)}\right)\label{eq:hp}
\end{equation}

\begin{equation}
24\beta K_{0}^{\prime}=\ln\left(\cosh\beta(3k+h)\cosh\beta(3k-h)\right)+\ln\left(\cosh\beta(k+h)\cosh\beta(k-h)\right)^{3}.\label{eq:k0}
\end{equation}

These relations are, in turn, used to correct the tensors $J,K$ in
App.$~\ref{sec:clauses}$. To do so, an inversion of the above relations
is needed which may yield multiple solution. See \ref{sec:invertibility}
for more details.

\subsection{\label{sec:invertibility}Procedure to invert the RG equations}

\begin{figure}[H]
\begin{centering}
\includegraphics[scale=0.6]{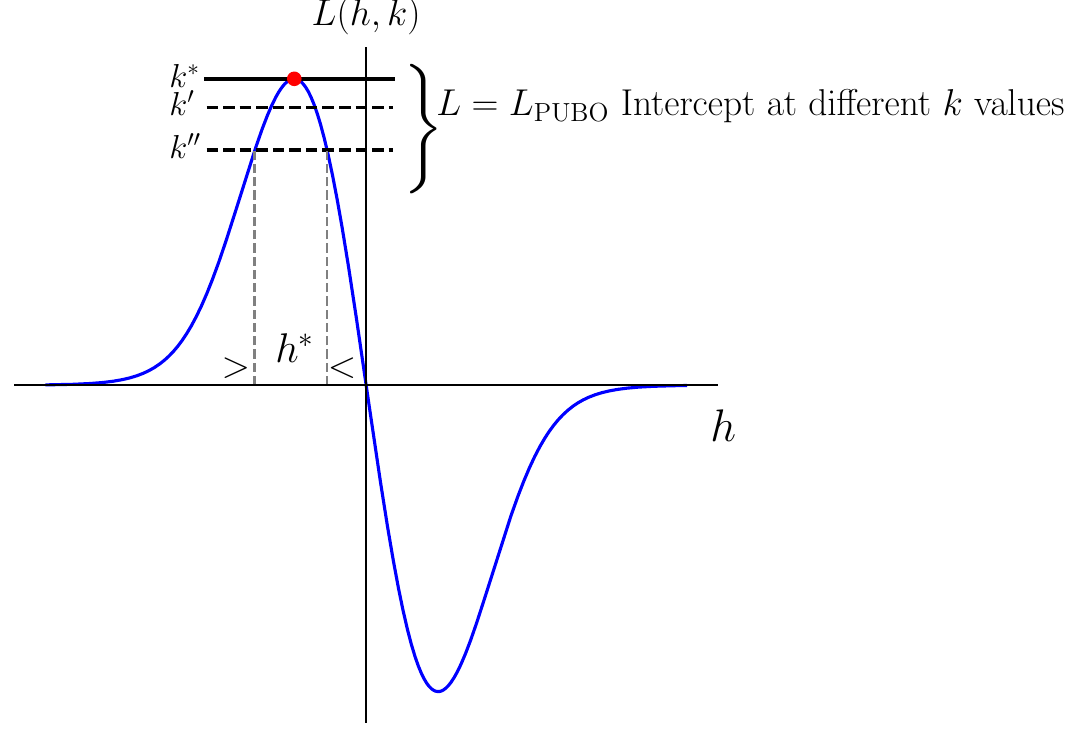}
\par\end{centering}
\caption{Graph of right hand side of Eq.$~\ref{eq:P-term}$, with $\beta=1$
and fixed $k$, as a function of $h$. The inversion is performed
by intersecting a horizontal line ($L$) with the given curve. For
optimal performance the parameter $k=k^{*}$ is chosen such that the
intersection is at the peak of the curve. The x-coordinate of the
intersection is the parameter $h=h^{*}$, which could have either
two, one or even zero possible values, depending on the value of $k$that
is chosen.\label{fig:L_invert}}
\end{figure}

To find $(h,k)$ as a function of a given value of $L$, we need to
invert Eq.$~\ref{eq:P-term}$, see Fig.$~\ref{fig:L_invert}$. The
parameter $k$ is an overall scaling of the curve $\left\{ L(h,k),h\right\} $
and is set so that the intersection with the horizontal line $L_{\mathrm{PUBO}}$
is unique. Then, for this value of $k=k^{*}(\beta)$, the x-intercept
gives the value of $h=h^{*}$, such that $L(k^{*},h^{*})=L_{\mathrm{PUBO}}$.

We observe that a choice of parameters that makes the correspondence
unique yields better performance for the QUBO-RG simulation, however
a non-unique correspondence is also possible. This bears no effect
with regard to circumventing the extra spins (\ref{sec:Effective-Interaction})
or ground state preservation (\ref{sec:Ground}), as any correspondence
preserves the partition function. By moving away from a unique solution,
the magnitude of the QUBO-RG local fields increases, thus explaining
the worsening of performance, and its energy levels become similar
to the Rosenberg energy levels, see table$~\ref{tab:compare_rosen}$.

\subsection{\label{sec:Invariant}Invariant parameters of the partial summation}

The partial summation in Eq.$~\ref{Hq}$ admits an invariant parameter
$\xi$, whenever it multiplies degrees of freedom that are not being
summed over ($s_{2}$ and $s_{3}$ below):

\begin{gather}
-\beta H=\tikzmark{sa}{\underbrace{ks_{2}s_{1}+hs_{1}}}+\tikzmark{sc}{\xi}s_{2}s_{3}\\[2em]
-\beta H'={K'_{0}}+{h's_{2}}+{\xi}s_{2}s_{3},\label{eq:h_prime}
\end{gather}
\begin{tikzpicture}[overlay, remember picture, shorten >=5pt, shorten <=0pt]
    \draw[arrows=->] 
    ( $ (pic cs:sa) +(25.5pt,-1.8ex) $ ) -- 
    ( $ (pic cs:sa) +(16pt,-7ex) $ );
    \draw[arrows=->]
    ( $ (pic cs:sa) +(25.5pt,-1.8ex) $ ) -- 
    ( $ (pic cs:sa) +(35pt,-7.5ex) $ );
    \draw[arrows=->] 
    ( $ (pic cs:sc) +(5pt,-1ex) $ ) -- 
    ( $ (pic cs:sc) +(5pt,-7.6ex) $ );
\end{tikzpicture}

where $\sum_{s_{1}}e^{-\beta H}=e^{-\beta H'}$, and $K_{0}^{\prime},h^{\prime}$
are given in App.$~\ref{sec:Star-triangle-transform}$. With this
property, we can engineer any term of the resulting Hamiltonian $H'$
by pre-canceling the undesired terms already in $H$ and further inserting
the desired terms, here $J$. Thus we are able to cancel the term
$h^{\prime}$ of Eq.$~\ref{eq:h_prime}$ in the same way as the terms
$(\textcolor{red}{J'},\textcolor{red}{K'})$ of Eq.$~\ref{eq:transform}$:

\begin{gather}
-\beta H=\underbrace{{\tikzmark{sb}{ks_{2}s_{1}+hs_{1}}+\tikzmark{sd}{(J-h')}s_{2}}}\\[2em]
-\beta H'={K'_{0}}+{Js_{2}}.
\end{gather}

\begin{tikzpicture}[overlay, remember picture, shorten >=5pt, shorten <=0pt]
    \draw[arrows=->] 
    ( $ (pic cs:sb) +(53.5pt,-2ex) $ ) -- 
    ( $ (pic cs:sb) +(40pt,-7ex) $ );
    \draw[arrows=->]
    ( $ (pic cs:sb) +(53.5pt,-2ex) $ ) -- 
    ( $ (pic cs:sb) +(62pt,-7.5ex) $ );
    
\end{tikzpicture}

\subsection{\label{sec:clauses}Pseudocode to quadratize arbitrary number of
clauses}

In this section we show the pseudocode to implement the transform
from Eq. \eqref{eq:transform} to Eq. \eqref{Hq}.

\begin{algorithm}[H] 
\caption*{Pseudocode to transform a 3-SAT clause problem into a pariwise interacting system} 
\begin{algorithmic}[1] 
\State \textbf{Input:} $M, J,K,L,\beta$  
\State \textbf{Output:} $\overline{J}^Q,\overline{K}^Q$, new constant term $K'_0$

\State Calculate the value $k(\beta) = k^*$ which makes the inversion unique with one $h_e = h^*$, by procedure in Fig.~\ref{fig:L_invert}\;

\State Initialize QUBO-RG arrays $\overline{J}^Q$ and  $\overline{K}^Q$ of size $N_Q \times 1$ and $N_Q \times N_Q$ resp. with the values of $J,K$ for corresponding indices. The remaining entries will be corrected and filled in the following loop. See Eq.~\ref{eq:parallel-1}.\;
\Statex\;
\Statex Initialize counter for sequential loading:\;
\State count $=0$\;
\Statex Initialize overall constant in the Hamiltonian, stemming from general RG transforms $K'_0$, as in Eq.~\ref{eq:stt}:\;
\State $K'_0$ = 0\;
\For{$(i,j,l)$ such that $L[i,j,l] \neq 0$}

	\State count $+=1$\;
\Statex\;
\Statex\hspace{1.4em}Retrieve external field on extra spins $h_e = h^*$, which can have one of two values according to sign of $L[i,j,l]$:\;
    \State $h_e =$ Calculate\_he($L[i,j,l]$) via Eq.~\ref{eq:P-term}\;
	\Statex\;
\Statex\hspace{1.4em}Calculate corrections, which are functions of $(h_e,k):$\;
	\State $K' =$ Calculate\_kp($\beta k$,$\beta h_e$) via Eq.~\ref{eq:kp}\;
	\State $h' =$ Calculate\_hp($\beta k$,$\beta h_e$) via Eq.~\ref{eq:hp}\;
	
	\Statex\;

\Statex\hspace{1.4em}Accumulate constant term that carries part of the free energy:\;
	\State $K'_0 +=  $New\_$K'_0$($\beta k$,$\beta h_e$) via Eq.~\ref{eq:k0}\;
	\Statex\;

\Statex\hspace{1.4em}Implement corrections:\;
	\State $\overline{J}^Q[i],\overline{J}^Q[j],\overline{J}^Q[l] -= h' $\; 
	\State $\overline{K}^Q[i,j],\overline{K}^Q[i,l],\overline{K}^Q[j,l] -= K' $\;
\Statex\;
\Statex\hspace{1.3em}Fill in remaining entries:\;
	\State $\overline{J}^Q[$N + count -1$] = h_e $\;

	\State $\overline{K}^Q[i$, N + count -1$] = k^*$\;
	\State $\overline{K}^Q[j$, N + count -1$] = k^*$\;
	\State $\overline{K}^Q[l$, N + count -1$] = k^*$\;

\EndFor 
\State \Return $\overline{J}^Q,\overline{K}^Q, K'_0 + (M/8)$ 
\label{alg:alg_quadratization}
\end{algorithmic} 
\end{algorithm}

In the pairwise formalism, the local field $h_{i}^{Q}$ that QUBO-RG
(Eq.$~\ref{Hq}$) computes reads
\begin{equation}
h_{i}^{Q}=\overline{{J_{i}^{Q}}}+\sum_{l=1}^{N+N_{L}}\overline{{K_{il}^{Q}}}s_{l},\label{eq:QUBO_RG_field}
\end{equation}

where $N_{L}$ is the number of non-zero $L$(third-order) terms,
thus it already includes the contribution from the extra spins, due
to the enlarged matrices $\overline{J^{Q}},\overline{K^{Q}}$.

In contrast, the form of the local field of the PUBO solver is
\begin{equation}
h_{i}^{P}=J_{i}+\sum_{j=1}^{N}K_{ij}s_{j}+\sum_{j,k=1}^{N}L_{ijk}s_{j}s_{k}.\label{eq:PUBO_field}
\end{equation}

As indicated in algorithm $\ref{alg:alg_quadratization}$, to accomodate
the additional spins $s_{e}$ inside the new quadratic matrices (the
values of $h,k$), they are included inside $J^{Q},K^{Q}$, attached
after the values of the old matrices, with total size needed given
by:

\begin{equation}
\begin{cases}
|\overline{J}^{Q}|=N+N_{L}\equiv N_{Q}\\
|\overline{K}^{Q}|=(N+N_{L})\times(N+N_{L}),
\end{cases}\label{eq:parallel-1}
\end{equation}

where $N_{L}$ is the number of non-zero $L$(third-order) terms.
Note that this is very close to the number of clauses $M\equiv\alpha N$,
with $\alpha$ being the clause density. Yet, as one can have the
following problem $F:$ 

\begin{equation}
(x_{1}\lor x_{2}\lor x_{3})\land...\land(\ensuremath{\overline{x}_{1}}\lor\ensuremath{\overline{x}_{2}}\lor\ensuremath{\overline{x}_{3}})\land...
\end{equation}

these two clauses give together a vanishing $L_{123}$ term in the
Hamiltonian (Eq.$~\ref{eq:H_normal}$), because they involve the same
literals with opposing sign (see Eq.$~\ref{eq:tensors}$), but nevertheless
count as two additional clauses for $M$.

In this work, we only use these augmented matrices to compute the
QUBO-RG points in Fig.$~\ref{fig:fraction_red}$.

\subsection{\label{sec:Energies-and-Probabilities}Energies and probabilities
for states in a star-triangle transform}

We wish to demonstrate in table$~\ref{tab:compare_rosen}$ that the
energy levels of our QUBO-RG resemble the energy levels of Rosenberg
when the parameters are chosen away from the point at vanishing slope
$(k^{*},h^{*})$ from section $\ref{sec:invertibility}$.

\begin{table}[h!] \centering \small  \begin{tabular}{ S[table-format=1.0,table-align-text-post=false]  S[table-format=1.3]  S[table-format=1.3]  S[table-format=2.3]  S[table-format=2.3]  S[table-format=1.3] } \toprule {Configuration} & {$E_{\mathrm{PUBO}}$} & {$E_{\mathrm{QUBO\text{-}Rosenberg}}$} & {$E_{\mathrm{QUBO-RG\text{-}k''(\beta)}}$} & {$E_{\mathrm{QUBO-RG\text{-}k^{*}(\beta)}}$} & {$Prob.(\text{QUBO-RG)}$} \\
\midrule {-1 -1 -1 -1} & 1 & 1 & 1.000 & 1.002 & 0.050 \\
{-1 -1 -1 1} & 1 & 1 & 7.500 & 7.110 & 0.000 \\
\midrule {-1 -1 1 -1} & 0 & 1 & 0.000 & 0.032 & 0.131 \\
{-1 -1 1 1} & 0 & 0 & 2.500 & 3.458 & 0.000 \\
\midrule {-1 1 -1 -1} & 0 & 0 & 0.000 & 0.032 & 0.131 \\
{-1 1 -1 1} & 0 & 2 & 2.500 & 3.458 & 0.004 \\
\midrule {-1 1 1 -1} & 0 & 0 & 1.500 & 0.389 & 0.092 \\
{-1 1 1 1} & 0 & 1 & 0.000 & 1.133 & 0.043 \\
\midrule {1 -1 -1 -1} & 0 & 0 & 0.000 & 0.032 & 0.131 \\
{1 -1 -1 1} & 0 & 2 & 2.500 & 3.458& 0.004 \\
\midrule {1 -1 1 -1} & 0 & 0 & 1.500 & 0.389 & 0.092 \\
{1 -1 1 1} & 0 & 1 & 0.000 & 1.133 & 0.043 \\
\midrule {1 1 -1 -1} & 0 & 0 & 1.500 & 0.389 & 0.092 \\
{1 1 -1 1} & 0 & 4 & 0.000 & 1.132& 0.043 \\
\midrule {1 1 1 -1} & 0 & 0 & 5.500 & 2.072 & 0.017 \\
{1 1 1 1} & 0 & 3 & 0.000 & 0.135 & 0.119 \\
\bottomrule \end{tabular}\caption{Complete set of configurations pertaining to clause $(x_1 \lor x_2 \lor x_3)$ with $\beta = 1$. Energies in the native PUBO setting and corresponding probabilities in different quadratizations: Rosenberg, QUBO-RG($k^{''}$) at a non-ideal coupling parameter, and QUBO-RG($k^{*}$) where the ideal coupling parameter is chosen according to procedure in \ref{sec:invertibility}.} \label{tab:compare_rosen} \end{table}

From this table one can verify the following relations, caused by
the preservation of the partition function and by the definition $p(s)=e^{-\beta H(s)}/Z$
and where $P,Q$ stand for PUBO and QUBO-RG respectively:

\begin{equation}
p^{P}(s_{b})=p^{Q}(s_{b,}s_{e}=+1)+p^{Q}(s_{b,}s_{e}=-1).
\end{equation}
And likewise for the respective free energies, where we define $E_{\pm}=E(s_{b},s_{e}=\pm1)$,
$p_{\pm}=p^{Q}(s_{b},s_{e}=\pm1)$, with $k_{B}=1$:

\begin{equation}
Ep^{P}+Tp^{P}\ln p^{P}=(E_{+}p_{+}+E_{-}p_{-})+T(p_{+}\ln p_{+}+p_{-}\ln p_{-})\,.\label{eq:free_en}
\end{equation}

\subsection{\label{sec:Comparison_QUBO_PUBO}Comparison between PUBO and QUBO-RG
local fields}

\begin{figure}[H]
\includegraphics[scale=0.5]{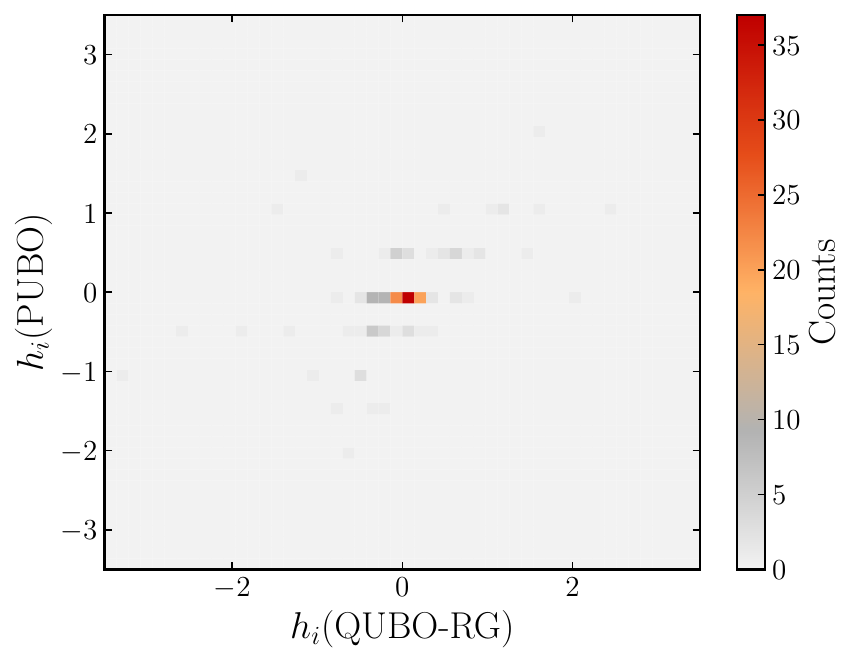}

\caption{QUBO-RG vs PUBO density plot of local fields $h_{i}$ of base spins.
One run at $\beta=5$ of duration $\tau_{s}=10^{5}$ steps was performed
on a 3-SAT problem of size $N=100$ and transformed with our quadratization.
At every update of a base spin, the QUBO-RG local field (Eq.$~\ref{eq:QUBO_RG_field}$)
is computed and shown on the x-axis. The local field that a PUBO solver
would compute from base spins alone is shown on the y-axis, at the
same time step.}
\label{fig:qubo_pubo_density}
\end{figure}

In Fig.$~\ref{fig:qubo_pubo_density}$ we show a density plot of the
alignment between the QUBO-RG fields and PUBO fields. The poor alignment
is a common effect of quadratization transformations which change
the local field, leading to poor performance. Due to mostly extra
spins being sampled and quick convergence toward vanishing local PUBO
field, such a plot admits a high concentration of points at vanishing
PUBO field, with relatively few point outside this range.

\begin{figure}[H]
\includegraphics[scale=0.6]{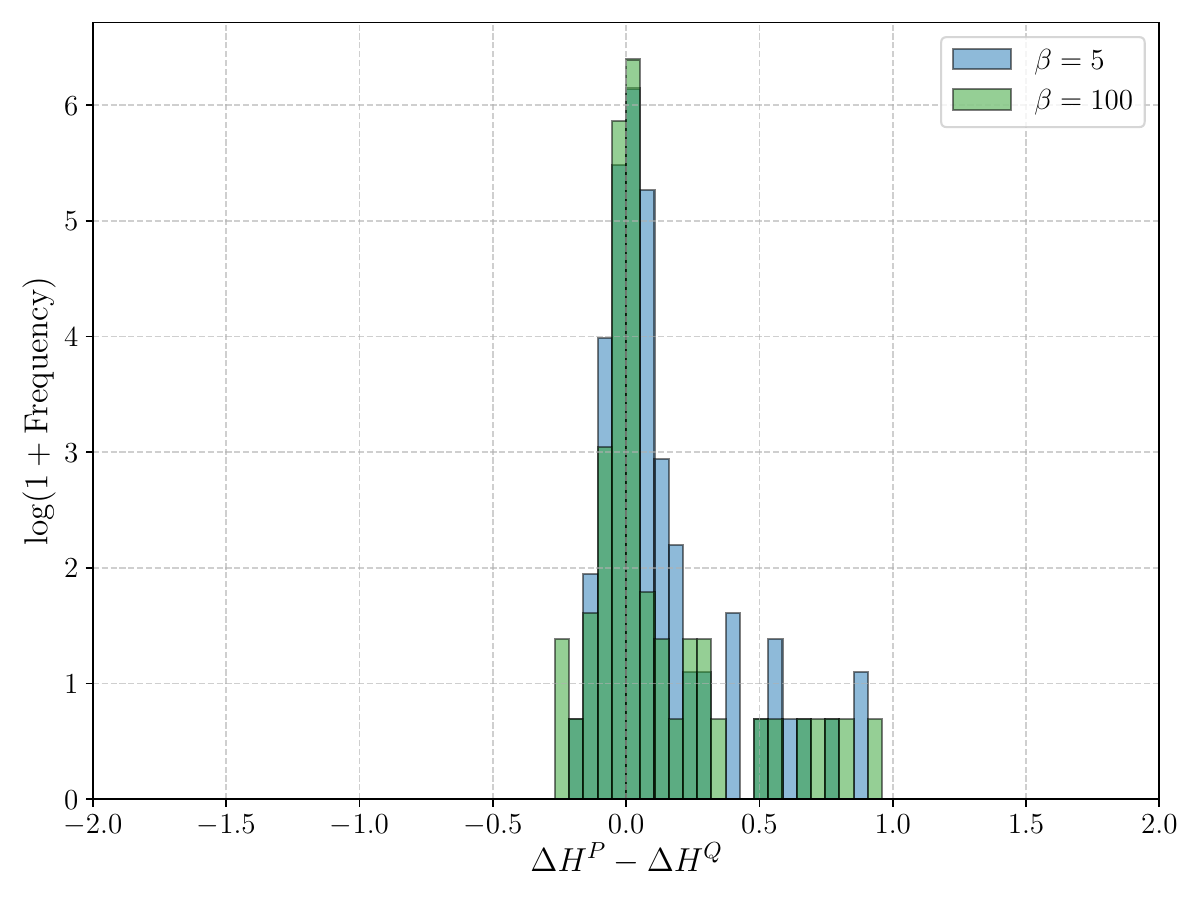}

\caption{\label{fig:PUBO-vs-QUBO}A QUBO-RG run is performed on a 3-SAT problem
of size $N=100$ for a total of $\tau_{s}=10^{7}$ steps. Upon a base
spin update, both a local QUBO-RG field (Eq.$~\ref{eq:QUBO_RG_field}$)
and local PUBO field (Eq.$~\ref{eq:PUBO_field}$) are calculated,
and the difference is plotted. This difference amounts to the difference
in energies caused by flipping the updated spin. A total of 100 runs
with random initial condition were performed for two different temperatures
$\beta$. The average value of the difference of local fields encountered
over the 100 runs is shown on the x-axis in order to encompass general
behavior of the solver, and the frequency of occurrences on the y-axis.}
\end{figure}

Another possible comparison is by plotting differences in local fields
of PUBO $h_{i}^{P}$ and QUBO-RG $h_{i}^{Q}$ (or, correspondingly,
$\Delta H^{P/Q}=-2h_{i}^{P/Q}$). We investigate these differences
at two temperatures. The figure \ref{fig:PUBO-vs-QUBO} shows that
these differences become smaller with higher $\beta$, indicating
that the two systems approach each other at small temperatures. 

\subsection{\label{sec:qubo_dynamics}QUBO-RG in the long runtime limit}

\begin{figure}[H]
\includegraphics[scale=0.6]{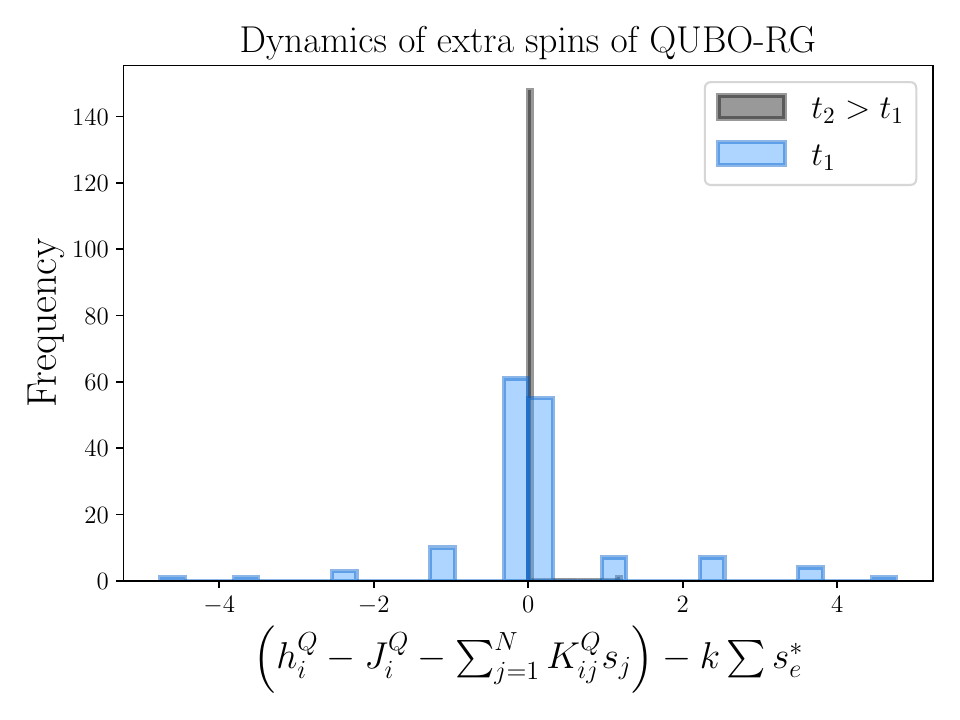}

\caption{A QUBO-RG run is performed on a 3-SAT problem of size $N=100$ at
$\beta=10$ for a total of $\tau_{s}=10^{7}$ steps. Upon update of
a base spin, its local field $h_{i}^{Q}$ is measured for 150 updates
$t<t_{1}$ and 150 updates for $t>t_{2}$; $t_{1}=10^{3}$ and $t_{2}=10^{4}$.
The effect of the base spins is subtracted $h_{i}^{Q}-J_{i}^{Q}-\sum_{j=1}^{N}K_{ij}^{Q}s_{j}$,
giving $k\sum_{e}s_{e}$. For $\beta\to\infty$ this value corresponds
to the contribution from the fictitious spins $k\sum_{e}s_{e}^{*}$
cf. Eq.$~\ref{eq:fict}$. The difference of these two terms is shown
on the x-axis and the frequency of the resulting value is plotted.
For $t>t_{2}$, the histogram concentrates around $0$, showing convergence
of the extra spins with time toward $s_{e}^{*}$. \label{fig:QUBO_long}}
\end{figure}

In Fig.$~\ref{fig:QUBO_long}$ we analyze the dynamics of QUBO-RG
in its long runtime limit. From the full QUBO-RG interaction with
base and extra spins (Eq.$~\ref{eq:QUBO_RG_field}$), we subtract
the effect of the base spins, thus yielding solely the contribution
of QUBO-RG extra spins $s_{e}$. The degree to which the extra spins
in a QUBO-RG simulation differ from the value given by Eq.$~\eqref{eq:parallel}$
of the QUBO-Tr($\infty$) system (here spins $s_{e}^{*}$) is measured
at two different time steps $t_{1}$ (early) and $t_{2}$ (late),
verifying eventual convergence of QUBO-RG towards QUBO-Tr($\infty$).
This implies that the two energy landscapes, traversed by QUBO-Tr($\infty$)
and QUBO-RG, are eventually the same landscape.

\subsection{\label{sec:pt_schedule}Parallel tempering schedule analysis}

We here follow mostly the analysis in ref.\cite{PhysRevApplied.17.024052},
although more optimized precedures exist \cite{chowdhury2025pushingboundaryquantumadvantage}.

\begin{figure}[H]
\includegraphics[scale=0.6]{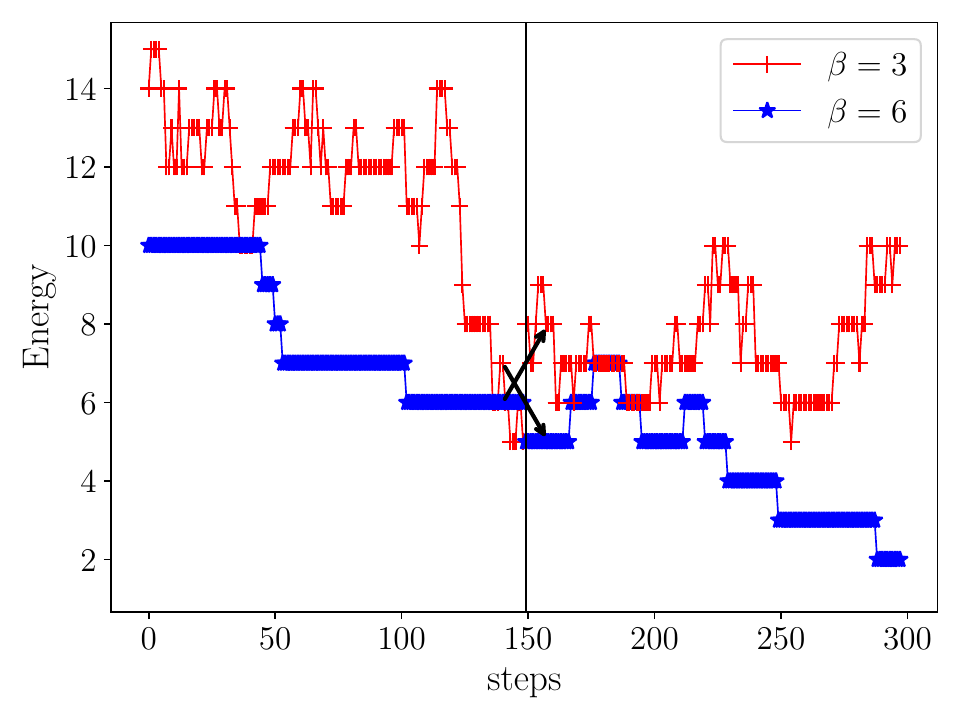}

\caption{Representative run of parallel tempering with temperature swap around
step 149 (vertical line). The initial condition at the new temperature
is copied from the final state of the old temperature, at step 149.\label{fig:rep_run}}
\end{figure}

In parallel tempering, replicas of the system run in parallel at different
temperatures. Higher temperatures serve, in heuristic terms, as the
diversification element of the algorithm: their highly stochastic
behavior and strong fluctuations allow them to explore a larger ensemble
of states in the search space, until they find a configuration with
a lower energy than the neighboring replicas at lower temperature.
At this iteration a switching occurs, the state held by the high temperature
replica is exchanged with that of the low temperature one. Because
of the low stochasticity at small temperatures, cold replicas are
usually unable to get out of deep local metastable minima, so they
require hotter replicas to find better starting states for them to
minimize. In Fig.$~\ref{fig:rep_run}$ we illustrate the implementation
of such a run, where the red (blue) line corresponds to high (low)
temperature.

The number of replicas and their temperatures are important parameters
to set. The ideal configuration has each replica exploring an ensemble
of states that slightly overlaps with those of the hotter and colder
replicas. We have used an empirical approach based on three main criteria:

\begin{enumerate} \item Fast computation and low memory occupancy. We have chosen as few replicas as possible and 1 swap step between runs. \item A replica explores the solution space of a problem with energy cost fluctuating around an average energy value and with a standard deviation that depends on the temperature value. The explored solution space for a given $i$ th replica should overlap with that of its neighbors ($i+1$)th and ($i - 1$)th replica. \item The schedule of $\beta$ is approximated to be constant independent of $N$, for fixed $\alpha$. We perform an optimization of resources at $N =100$. \end{enumerate}

We hypothesize that for very high $\beta$ (zero-temperature limit),
parallel tempering does not offer additional speedup besides effectively
serving as a different initial condition for the solver. This is,
we assume that high absolute values of $\beta$ do not contribute
to an increase of performance significantly. We therefore employ more
replicas at low $\beta$, allowing us to have a finer spacing $\Delta\beta$
between $\beta's$ in that regime. In table$~\ref{tab:schedule}$
we analyze the performance while varying the spacing $\Delta\beta$
of the schedule, by increasing the maximum $\beta$ allowed while
assuming a constant number of 20 replicas throughout.

\begin{table}[H]
\begin{table}[H]     \centering         \label{tab:schedule}     \vspace{4pt}     \begin{tabular}{@{\hspace{4pt}} c c @{\hspace{4pt}}}         \toprule         \textbf{UTS-99} ($\times 10^4$) & \textbf{$\beta_{\max}$ (20 replicas)} \\         \midrule         $\infty$ & $2$ \\         $9.1$ & $5$ \\         $2.8$ & $10$ \\         $2.2$ & $20$ \\         $2.1$ & $40$ \\         $1.9$ & $60$ \\         $2.3$ & $80$ \\         \bottomrule     \end{tabular} \end{table}

\caption{Inverse temperature schedule for 20 replicas. The iteration count
($\mathrm{UTS}_{99}$) is given in units of $10^{4}$ and the performance
vs temperature schedule spacing is evaluated for an $N=100$ 3-SAT
problem during a PUBO run.}
\label{tab:schedule}
\end{table}

\begin{figure}[H]
\includegraphics[scale=0.6]{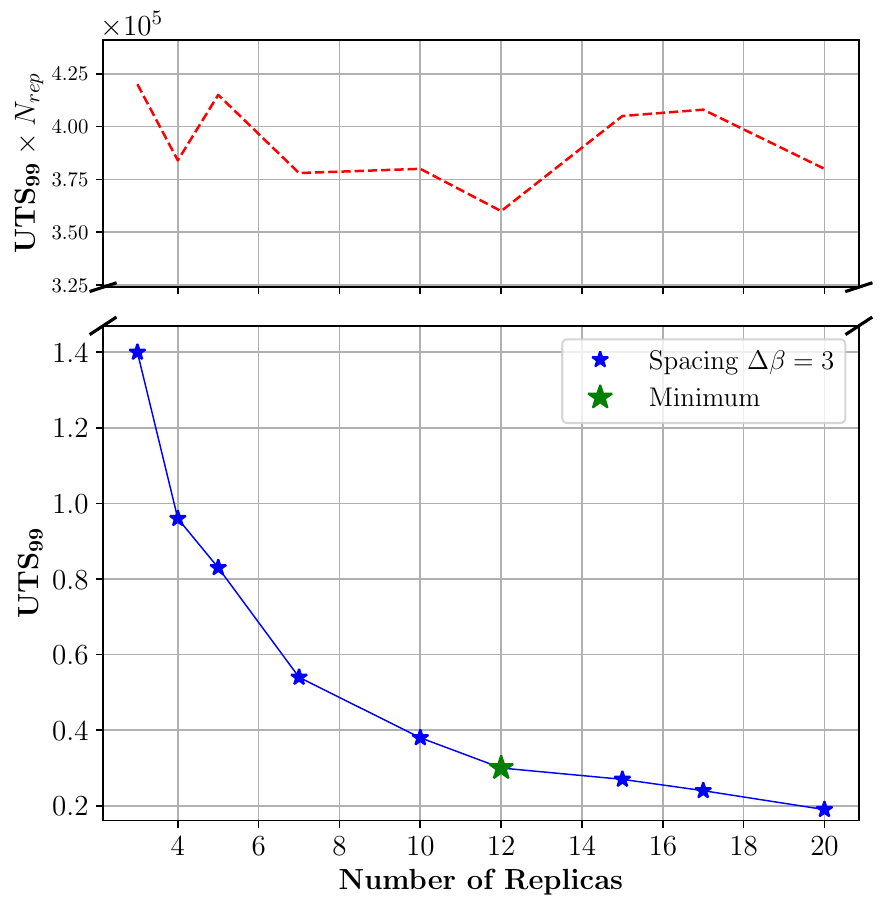}

\caption{Lower curve: We plot the performance for different numbers of replicas
at the fixed spacing $\Delta\beta=3$. Each point corresponds to an
average of 100 solved instances where each was attempted 100 times
with random initial conditions. Upper curve: the data form UTS is
combined with the number of replicas, yielding a minimum number of
12.}
\label{fig:rep_number}
\end{figure}

We find an optimum point for $\beta_{\mathrm{max}}=60$, which means
a spacing of $\Delta\beta=3$ . Once $\Delta\beta$ is fixed, we vary
the number of replicas and again plot the measure $\mathrm{UTS}_{99}$(roughly
the number of spin updates that guarantees 99 instances solved out
of 100) as given in Fig.$~\ref{fig:rep_number}$. As we aim for resource
efficiency, we pick a number of replicas which allows for a good trade-off
between performance and energy. The cost is proportional to $(\mathrm{UTS}\times\text{number of replicas)}$.
We find that we have a minimum for $12$ replicas. Therefore, we pick
this schedule to do the analysis of our algorithms.

\begin{figure}[H]
\includegraphics[scale=0.6]{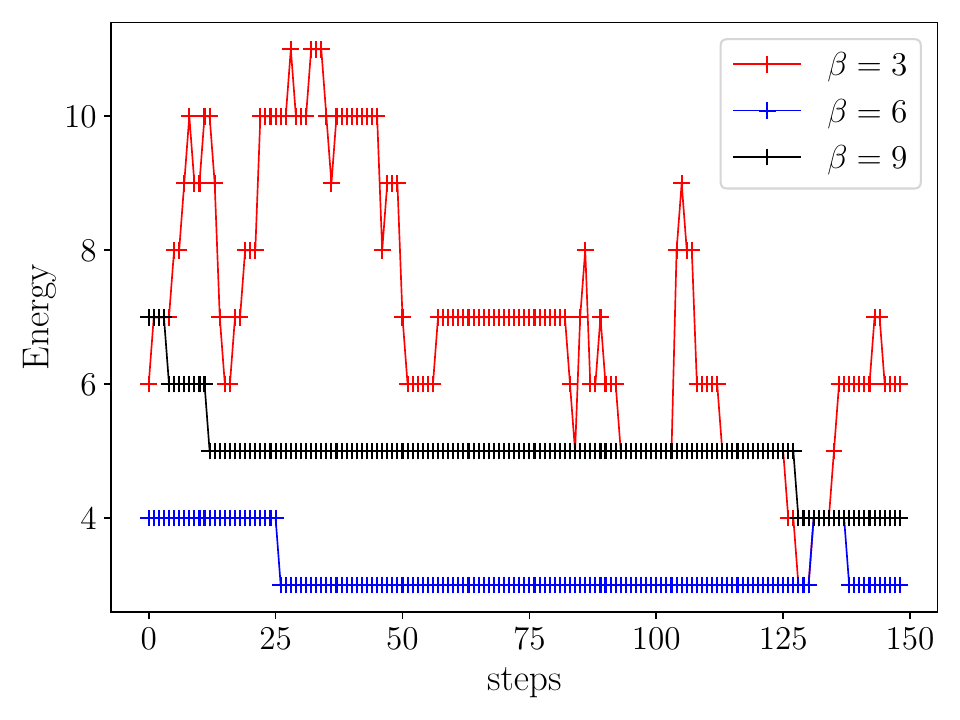}

\caption{A PUBO run is performed for an $N=100$ problem, for three different
replicas in parallel. Plotted is the energy over time that each replica
visits.}
\label{fig:overlap}
\end{figure}

Finally, adjacent replicas must share energy values, satisfying criterion
2 above. We plot in Fig.$~\ref{fig:overlap}$ the energy as a function
of steps of three runs at different temperatures, showing their overlap
in energy.

\subsection{\label{sec:Alt_alg}Alternative Algorithms}

\subsubsection{\label{subsec:Signum-approach}Using a Sign function to compute the
extra spin contribution}

In order to facilitate a potential hardware implementation of QUBO-Tr
described by Eq.$~\ref{eq:dissect}$, we here show an alternative
to the logcosh function. As the logcosh depends on the discrete variable
$s_{j}+s_{k}$, it is possible to express it using a simpler nonlinear
function, namely a $\operatorname{sign}$. Other simple nonlinear
functions may also accommodate this term, however we choose the $\operatorname{sign}$
due to the local field resemblance to both the $\beta\rightarrow\infty$
limit $\ref{eq:fict}$ and due to its binary output mirroring the
values of a spin variable, as follows:

\begin{equation}
\frac{1}{2\beta}\ln\frac{\cosh{\beta(k(s_{j}+s_{k})+h_{e}+k)}}{\cosh{\beta(k(s_{j}+s_{k})+h_{e}-k)}}=\zeta_{1}(s_{j}+s_{k})+\zeta_{2}+\delta\left[\operatorname{sgn}(k(s_{j}+s_{k})+h_{e})-\frac{1}{2}(s_{j}+s_{k})\right].\label{eq:signum}
\end{equation}
Parameters $\zeta_{1},\zeta_{2}$ are obtained, for every $\beta$,
from the condition that the expression needs to agree on the end-points
$s_{j}+s_{k}=\pm2$, which yields a linear equation. For these points,
$\operatorname{sgn}(k(s_{j}+s_{k})+h_{e})=\pm2$ respectively, thus
cancelling the square brackets. For $s_{j}+s_{k}=0$, the function
$\operatorname{sgn}(h_{e})=\pm1$ depending on the sign of $h_{e}$.
The parameter $\delta$ is then calculated to match the remaining
case, in which $s_{j}+s_{k}=0$, as follows:

\begin{equation}
\delta=\frac{1}{2\beta}\ln\frac{\cosh{\beta(h_{e}+k)}}{\cosh{\beta(h_{e}-k)}}-\zeta_{2}.\label{eq:sign_delta}
\end{equation}

There is no additional cost to the linear terms, since they may be
incorporated in the $\overline{J}^{Q},\overline{K}^{Q}$ matrices
at the start of the run, resulting in the local field $\tilde{{h_{i}^{Q}}}$.
The value of the $\operatorname{sign}$ may be encoded, as in section
\ref{subsec:Algorithm-2--}, in the value of an extra spin which is
set according to its local cavity field:

\begin{equation}
h_{i}^{P}=\tilde{{h_{i}^{Q}}}(s_{b})+\delta\sum_{e}\operatorname{sgn}(k(s_{j}+s_{k})+h_{e}).\label{eq:sign_field}
\end{equation}

Equation$~\ref{eq:sign_field}$ has the same form as the limit $\beta\to\infty$
\eqref{eq:fict}, albeit with different parameters. This implies that
QUBO-Tr at finite $\beta$ can be computed with the same algorithm
as the $\beta\rightarrow\infty$ limit.

A remark about the ruggedness of the landscape is in order. Researchers
\cite{doi:10.1126/science.aab3326} have shown that any Ising Hamiltonian
can be transformed into only pairwise, and even nearest-neighbor,
interactions while maintaining the partition function, thus preserving
global minima for instance. This means that a linear function for
the local field is always possible. From the Markov-chain algorithm
perspective though, we find here that some sort of nonlinear function
in the local field must be employed, in order to allow for fast descent
toward the global minimum of the landscape.

\subsubsection{\label{subsec:Look-up-table-approach}Look-up table approach}

Here we give an implementation of Eq.$~\ref{eq:dissect}$ that uses
very small look-up tables. The number of operations is, however, difficult
to estimate because of the necessity to retrieve non-zero elements
from $\boldsymbol{\mathbb{\mathbb{G}}}[\colon,i]$. We remark that
$\boldsymbol{\mathbb{\mathbb{G}}}$ is a sparse matrix with exactly
$3$ non-zero elements in every line (the indices of the base spins
connected to the extra spin corresponding to the line number), and
on average $3\alpha$ non-zero elements in every column.

\begin{algorithm}[H] 
\caption{QUBO-Tr: Modified Glauber dynamics with look-up table corrections to local fields} 
\begin{algorithmic}[1] 
\Statex Construction of matrix $\boldsymbol{\mathbb{\mathbb{G}}}$:\; 
\For {$ i\in \{0,...,N_Q-N-1\}$ and $ j\in \{0,...,N-1\}$} 
	\State $\boldsymbol{\mathbb{G}}[i,j] = 1_{(\bar{h}_{e(i)}^Q \text{ depends on } s_j)}$ 
\EndFor\;
\Statex\;
\Statex Construction of look-up tables:
\For {$ a\in \{0,1\}$ and $ b\in \{0,1,2\}$} 
	\State $\mathbb{LUP}[a,b] = \frac{1}{2\beta}\log\cosh \beta(2(a-1)k + |h_e|(2b-1) + k)$
	\State $\mathbb{LUM}[a,b] = \frac{1}{2\beta}\log\cosh \beta(2(a-1)k + |h_e|(2b-1) - k)$ 
\EndFor
\Statex\rule{\linewidth}{0.4pt}\;
\Statex\textbf{Monte Carlo}\;         
\State \textbf{Input:} $\boldsymbol{s}$,$J^Q,K^Q,\boldsymbol{\mathbb{\mathbb{G}}}, T_{sim}$, Array of external fields $h_e$\;
\For{$t \leq T_{sim}$}\;

	\State Pick one spin $s_i$ at random among $N$ possibles ones.\;
	
	\State $h_i =J^Q[i] + K^Q[i,\colon]\cdot \bold{s}$\;   
	\State $\Gamma$ $= \boldsymbol{\mathbb{\mathbb{G}}}\cdot\bold{s}$ (Array of $(s_i + s_j +s_k)$ for all extras, Eq.~\ref{eq:M})\;
	\State $S_e =\{\text{indices of non-zero entries of } \boldsymbol{\mathbb{\mathbb{G}}}[\colon ,i]\}$\;   
	\For{extra in $S_e$}\;          
		\State First index:  he\_ind = (sgn($h_e$[extra]) + 1)/2\;     
		\State Second index:  k\_ind $= 1 + (\Gamma[extra] - s_i)/2$\;     
		\State $h_i = h_i + \mathbb{LUP}[\text{he\_ind,k\_ind}] - \mathbb{LUM}[\text{he\_ind,k\_ind}]$\;   
	\EndFor 
	\State Update selected spin $s_i$ based on $h_i$ (Eq.~\ref{eq:Glauber})\;
	\State Increment t\;
\EndFor 	

\end{algorithmic} 
\end{algorithm}

\subsubsection{\label{subsec:Diagonal-matrix-approach}Diagonal matrix approach
to eliminate unnecessary local field calculations}

In Algorithm \ref{alg:1-count} we give a detailed accounting of the
number of operations from Algorithm \ref{alg:1} per each spin update,
where line 10 is decomposed into simpler lines and more efficient
computations, and highlighted in red. Relative to Algorithm \ref{alg:1},
we introduce the object $\text{diag}(\boldsymbol{\mathbb{\mathbb{G}}}[\colon,i])$
-- which is a diagonal matrix with diagonal elements equal to those
of $\boldsymbol{\mathbb{\mathbb{G}}}[\colon,i]$ -- in order to select
only the fields necessary for an update, instead of computing all
fields as in the main text.

Each line in the resulting matrix $\text{diag}(\boldsymbol{\mathbb{\mathbb{G}}}[\colon,i])\boldsymbol{\mathbb{\mathbb{G}}}$
is built by the dot product between one line of $\text{diag}(\boldsymbol{\mathbb{\mathbb{G}}}[\colon,i])$
with successive columns of $\boldsymbol{\mathbb{\mathbb{G}}}$. As
the line in $\text{diag}(\boldsymbol{\mathbb{\mathbb{G}}}[\colon,i])$
has (at most) one non-zero entry, this results in solely 3 products
to build one line of $\text{diag}(\boldsymbol{\mathbb{\mathbb{G}}}[\colon,i])\boldsymbol{\mathbb{\mathbb{G}}}$.
There are only $3\alpha$ lines of $\text{diag}(\boldsymbol{\mathbb{\mathbb{G}}}[\colon,i])$
which are non-zero, thus the total number of multiplications in this
matrix-matrix product in on average $9\alpha$.

\begin{algorithm}[H] \caption{QUBO-Tr: Modified Glauber Dynamics with lncosh on arrays and efficient implemention} \begin{algorithmic}[1]
\Statex \textbf{Construction of matrix} $\boldsymbol{\mathbb{G}}$: \For{$i \in \{0,\dots,N_Q - N - 1\}$}     \For{$j \in \{0,\dots,N - 1\}$}         \State $\boldsymbol{\mathbb{G}}[i,j] = 1_{(\bar{h}_{e(i)}^Q \text{ depends on } s_j)}$     \EndFor \EndFor
\Statex
\Statex \hrulefill \Statex \textbf{Monte Carlo} \State \textbf{Input:} $\boldsymbol{s}$, $J^Q$, $K^Q$, $\boldsymbol{\mathbb{G}}$, $T_{\text{sim}}$, array of external fields $h_e$ 
\For{$t \leq T_{\text{sim}}$}     \State Pick one spin $s_i$ at random among $N$ possible ones.     
\State $h_i \gets J^Q[i] + K^Q[i,:] \cdot \boldsymbol{s}$ \Comment{$6\alpha$ Additions + $6\alpha$ Multiplications}     
\State $\textcolor{red}{\chi  \gets \text{diag}(\mathbb{G}[:,i])\boldsymbol{\mathbb{G}}}$ \Comment{$3 \times 3\alpha$ Multiplications}     
\State $\textcolor{red}{h_e^Q \gets \chi \cdot \boldsymbol{s}}$ \Comment{$6\alpha$ Additions + $9\alpha$ Multiplications}     
\State \textcolor{red}{$h_e^Q \gets h_e^Q + \text{diag}(\mathbb{G}[:,i])h_e$} \Comment{ $3\alpha$ Additions + $3\alpha$ Multiplications}     \State $h_e^{Q+} \gets h_e^Q + \boldsymbol{\mathbb{G}}[:, i](k - k s_i)$ \Comment{ $3\alpha$ Additions + $(3\alpha)$ Multiplications}     \State $h_e^{Q-} \gets h_e^Q + \boldsymbol{\mathbb{G}}[:, i](-k - k s_i)$ \Comment{ $3\alpha$ Additions + $(3\alpha)$ Multiplications}     \State $h_i \gets h_i + \boldsymbol{\mathbb{G}}[:, i]^T \cdot \dfrac{1}{2\beta} \left( \ln \cosh \beta h_e^{Q+} - \ln \cosh \beta h_e^{Q-} \right)$ \Comment{ $(6\alpha)$ log + $(6\alpha)$ Additions + $(9\alpha)$ Multiplications}     \State Update selected spin $s_i$     \State Increment $t$ \EndFor
\end{algorithmic} \label{alg:1-count} \end{algorithm}

For the PUBO case, in the best possible scenario there are $9\alpha$
operations: there are on average $3\alpha$ non-zero $L_{ijk}$ terms,
and 3 operations for every non-zero $L_{ijk}$ (Eq. $~\ref{eq:L_count}$):

\begin{equation}     h_i =h_i + L_{ijk}s_js_k \implies \text{2 Multiplications + 1 Addition }. \label{eq:L_count}\end{equation}

The total counting for the above algorithm is summarized in the following
table$~\ref{tab:counting}$, and compared to a PUBO update. This estimate
shows that no hidden exponential computations appear in the correction
term.\begin{table}[h!] \centering \renewcommand{\arraystretch}{1.3} \begin{tabular}{l c c c} \toprule \textbf{Type of operation} & \textbf{PUBO} & \textbf{QUBO-Tr} & \textbf{QUBO-RG / Rosen.} \\ \midrule Addition & $9 \alpha$ & $27 \alpha$ & $9 \alpha$ \\ Multiplication & $12 \alpha$ & $39 \alpha$ & $9 \alpha$ \\ Element-wise logcosh (log) & $0$ & $6 \alpha$ & $0$ \\ \midrule \textbf{Total} & $21 \alpha$ & $72 \alpha$ & $18 \alpha$ \\ \bottomrule \end{tabular} \caption{Total number of operations for the different approaches studied. Here we assume all types of operations have equal cost, namely additions, multiplications, logcosh, sign and engineering of diagonal matrices from column matrices. } \label{tab:counting} \end{table}  

\begin{thebibliography}{85}
\expandafter\ifx\csname natexlab\endcsname\relax\def\natexlab#1{#1}\fi
\expandafter\ifx\csname bibnamefont\endcsname\relax
  \def\bibnamefont#1{#1}\fi
\expandafter\ifx\csname bibfnamefont\endcsname\relax
  \def\bibfnamefont#1{#1}\fi
\expandafter\ifx\csname citenamefont\endcsname\relax
  \def\citenamefont#1{#1}\fi
\expandafter\ifx\csname url\endcsname\relax
  \def\url#1{\texttt{#1}}\fi
\expandafter\ifx\csname urlprefix\endcsname\relax\def\urlprefix{URL }\fi
\providecommand{\bibinfo}[2]{#2}
\providecommand{\eprint}[2][]{\url{#2}}

\bibitem[{\citenamefont{Kautz and Selman}(1992)}]{Kautz1992}
\bibinfo{author}{\bibfnamefont{H.~A.} \bibnamefont{Kautz}} \bibnamefont{and}
  \bibinfo{author}{\bibfnamefont{B.}~\bibnamefont{Selman}}, in
  \emph{\bibinfo{booktitle}{Proceedings of the 10th European Conference on
  Artificial Intelligence (ECAI)}} (\bibinfo{year}{1992}), pp.
  \bibinfo{pages}{359--363}.

\bibitem[{\citenamefont{Smith et~al.}(2008)\citenamefont{Smith, Frank, and
  Cushing}}]{Smith2008}
\bibinfo{author}{\bibfnamefont{D.~E.} \bibnamefont{Smith}},
  \bibinfo{author}{\bibfnamefont{J.}~\bibnamefont{Frank}}, \bibnamefont{and}
  \bibinfo{author}{\bibfnamefont{W.}~\bibnamefont{Cushing}},
  \bibinfo{journal}{AI Magazine} \textbf{\bibinfo{volume}{29}},
  \bibinfo{pages}{61} (\bibinfo{year}{2008}).

\bibitem[{\citenamefont{Biere et~al.}(1999)\citenamefont{Biere, Cimatti,
  Clarke, and Zhu}}]{Biere2003}
\bibinfo{author}{\bibfnamefont{A.}~\bibnamefont{Biere}},
  \bibinfo{author}{\bibfnamefont{A.}~\bibnamefont{Cimatti}},
  \bibinfo{author}{\bibfnamefont{E.}~\bibnamefont{Clarke}}, \bibnamefont{and}
  \bibinfo{author}{\bibfnamefont{Y.}~\bibnamefont{Zhu}}, in
  \emph{\bibinfo{booktitle}{Proceedings of the 5th International Conference on
  Tools and Algorithms for the Construction and Analysis of Systems (TACAS)}}
  (\bibinfo{publisher}{Springer}, \bibinfo{year}{1999}), pp.
  \bibinfo{pages}{193--207}.

\bibitem[{\citenamefont{Moskewicz et~al.}(2001)\citenamefont{Moskewicz,
  Madigan, Zhao, Zhang, and Malik}}]{Moskewicz2001}
\bibinfo{author}{\bibfnamefont{M.~W.} \bibnamefont{Moskewicz}},
  \bibinfo{author}{\bibfnamefont{C.~F.} \bibnamefont{Madigan}},
  \bibinfo{author}{\bibfnamefont{Y.}~\bibnamefont{Zhao}},
  \bibinfo{author}{\bibfnamefont{L.}~\bibnamefont{Zhang}}, \bibnamefont{and}
  \bibinfo{author}{\bibfnamefont{S.}~\bibnamefont{Malik}},
  \bibinfo{journal}{Proceedings of the 38th Annual Design Automation Conference
  (DAC)} pp. \bibinfo{pages}{530--535} (\bibinfo{year}{2001}).

\bibitem[{\citenamefont{Stojadinovi{\'{c}}}(2014)}]{Stojadinovic2014}
\bibinfo{author}{\bibfnamefont{M.}~\bibnamefont{Stojadinovi{\'{c}}}}, in
  \emph{\bibinfo{booktitle}{Principles and Practice of Constraint
  Programming}}, edited by
  \bibinfo{editor}{\bibfnamefont{B.}~\bibnamefont{O'Sullivan}}
  (\bibinfo{publisher}{Springer International Publishing},
  \bibinfo{address}{Cham}, \bibinfo{year}{2014}), pp.
  \bibinfo{pages}{886--902},
  \urlprefix\url{https://doi.org/10.1007/978-3-319-10428-7_63}.

\bibitem[{\citenamefont{Williamson et~al.}(1997)\citenamefont{Williamson, Hall,
  Hoogeveen, Hurkens, Lenstra, Sevast'janov, and
  Shmoys}}]{doi:10.1287/opre.45.2.288}
\bibinfo{author}{\bibfnamefont{D.~P.} \bibnamefont{Williamson}},
  \bibinfo{author}{\bibfnamefont{L.~A.} \bibnamefont{Hall}},
  \bibinfo{author}{\bibfnamefont{J.~A.} \bibnamefont{Hoogeveen}},
  \bibinfo{author}{\bibfnamefont{C.~A.~J.} \bibnamefont{Hurkens}},
  \bibinfo{author}{\bibfnamefont{J.~K.} \bibnamefont{Lenstra}},
  \bibinfo{author}{\bibfnamefont{S.~V.} \bibnamefont{Sevast'janov}},
  \bibnamefont{and} \bibinfo{author}{\bibfnamefont{D.~B.}
  \bibnamefont{Shmoys}}, \bibinfo{journal}{Operations Research}
  \textbf{\bibinfo{volume}{45}}, \bibinfo{pages}{288} (\bibinfo{year}{1997}),
  \eprint{https://doi.org/10.1287/opre.45.2.288},
  \urlprefix\url{https://doi.org/10.1287/opre.45.2.288}.

\bibitem[{\citenamefont{Aceto et~al.}(2004)\citenamefont{Aceto, Hansen, {Anna
  Ing{\'o}lfsd{\'o}ttir}, Johnsen, and Knudsen}}]{Aceto2004}
\bibinfo{author}{\bibfnamefont{L.}~\bibnamefont{Aceto}},
  \bibinfo{author}{\bibfnamefont{J.~A.} \bibnamefont{Hansen}},
  \bibinfo{author}{\bibnamefont{{Anna Ing{\'o}lfsd{\'o}ttir}}},
  \bibinfo{author}{\bibfnamefont{J.}~\bibnamefont{Johnsen}}, \bibnamefont{and}
  \bibinfo{author}{\bibfnamefont{J.}~\bibnamefont{Knudsen}},
  \bibinfo{journal}{Journal of Computer Science and Technology}
  \textbf{\bibinfo{volume}{19}}, \bibinfo{pages}{42} (\bibinfo{year}{2004}),
  \urlprefix\url{https://doi.org/10.1007/BF02944784}.

\bibitem[{\citenamefont{Jiang et~al.}(2002)\citenamefont{Jiang, Lin, Ma, and
  Zhang}}]{Jiang2002}
\bibinfo{author}{\bibfnamefont{T.}~\bibnamefont{Jiang}},
  \bibinfo{author}{\bibfnamefont{G.}~\bibnamefont{Lin}},
  \bibinfo{author}{\bibfnamefont{B.}~\bibnamefont{Ma}}, \bibnamefont{and}
  \bibinfo{author}{\bibfnamefont{K.}~\bibnamefont{Zhang}}, in
  \emph{\bibinfo{booktitle}{Proceedings of the 6th Annual International
  Conference on Computational Biology (RECOMB)}} (\bibinfo{publisher}{ACM},
  \bibinfo{year}{2002}), pp. \bibinfo{pages}{219--226}, \bibinfo{note}{uses SAT
  encodings to find minimal-energy conformations in protein folding}.

\bibitem[{\citenamefont{Dilkina et~al.}(2007)\citenamefont{Dilkina, Gomes, and
  Sabharwal}}]{Dilkina2007}
\bibinfo{author}{\bibfnamefont{B.}~\bibnamefont{Dilkina}},
  \bibinfo{author}{\bibfnamefont{C.~P.} \bibnamefont{Gomes}}, \bibnamefont{and}
  \bibinfo{author}{\bibfnamefont{D.}~\bibnamefont{Sabharwal}},
  \bibinfo{journal}{Constraints} \textbf{\bibinfo{volume}{12}},
  \bibinfo{pages}{375} (\bibinfo{year}{2007}), \bibinfo{note}{demonstrates
  SAT-based search for biological network and folding problems}.

\bibitem[{\citenamefont{{Anaconda Inc.}}(2023)}]{AnacondaCondaPerformance2025}
\bibinfo{author}{\bibnamefont{{Anaconda Inc.}}},
  \emph{\bibinfo{title}{Understanding and improving {Conda}'s performance}},
  \bibinfo{howpublished}{\url{https://www.anaconda.com/blog/understanding-and-improving-condas-performance}}
  (\bibinfo{year}{2023}), \bibinfo{note}{accessed: May 13, 2025}.

\bibitem[{\citenamefont{Ramamoorthy and Jayagowri}(2023)}]{RAMAMOORTHY20232539}
\bibinfo{author}{\bibfnamefont{A.}~\bibnamefont{Ramamoorthy}} \bibnamefont{and}
  \bibinfo{author}{\bibfnamefont{P.}~\bibnamefont{Jayagowri}},
  \bibinfo{journal}{Materials Today: Proceedings}
  \textbf{\bibinfo{volume}{80}}, \bibinfo{pages}{2539} (\bibinfo{year}{2023}),
  ISSN \bibinfo{issn}{2214-7853}, \bibinfo{note}{sI:5 NANO 2021},
  \urlprefix\url{https://www.sciencedirect.com/science/article/pii/S2214785321047982}.

\bibitem[{\citenamefont{Nohl et~al.}(2008)\citenamefont{Nohl, Malan, Starbug,
  and Pl\"otz}}]{Nohl2008}
\bibinfo{author}{\bibfnamefont{K.}~\bibnamefont{Nohl}},
  \bibinfo{author}{\bibfnamefont{D.}~\bibnamefont{Malan}},
  \bibinfo{author}{\bibnamefont{Starbug}}, \bibnamefont{and}
  \bibinfo{author}{\bibfnamefont{H.}~\bibnamefont{Pl\"otz}}, in
  \emph{\bibinfo{booktitle}{USENIX Security Symposium}} (\bibinfo{year}{2008}),
  pp. \bibinfo{pages}{185--193}, \bibinfo{note}{methodology related to GSM/A5/1
  key recovery via SAT-based cryptanalysis}.

\bibitem[{\citenamefont{Barkan et~al.}(2008)\citenamefont{Barkan, Biham, and
  Keller}}]{Barkan2003}
\bibinfo{author}{\bibfnamefont{E.}~\bibnamefont{Barkan}},
  \bibinfo{author}{\bibfnamefont{E.}~\bibnamefont{Biham}}, \bibnamefont{and}
  \bibinfo{author}{\bibfnamefont{N.}~\bibnamefont{Keller}},
  \bibinfo{journal}{Journal of Cryptology} \textbf{\bibinfo{volume}{21}},
  \bibinfo{pages}{392} (\bibinfo{year}{2008}), \bibinfo{note}{cryptanalysis
  approach enabling SAT reductions for GSM A5/1}.

\bibitem[{\citenamefont{Mironov and Zhang}(2006)}]{cryptoeprint:2006/254}
\bibinfo{author}{\bibfnamefont{I.}~\bibnamefont{Mironov}} \bibnamefont{and}
  \bibinfo{author}{\bibfnamefont{L.}~\bibnamefont{Zhang}},
  \emph{\bibinfo{title}{Applications of {SAT} solvers to cryptanalysis of hash
  functions}}, \bibinfo{howpublished}{Cryptology {ePrint} Archive, Paper
  2006/254} (\bibinfo{year}{2006}),
  \urlprefix\url{https://eprint.iacr.org/2006/254}.

\bibitem[{\citenamefont{Heule et~al.}(2016)\citenamefont{Heule, Kullmann, and
  Marek}}]{Heule2016}
\bibinfo{author}{\bibfnamefont{M.~J.~H.} \bibnamefont{Heule}},
  \bibinfo{author}{\bibfnamefont{O.}~\bibnamefont{Kullmann}}, \bibnamefont{and}
  \bibinfo{author}{\bibfnamefont{V.~W.} \bibnamefont{Marek}},
  \bibinfo{journal}{Communications of the {ACM}} \textbf{\bibinfo{volume}{59}},
  \bibinfo{pages}{107} (\bibinfo{year}{2016}), \bibinfo{note}{demonstrates
  large-scale SAT solving in mathematics; methods related to Erd{\H{o}}s
  discrepancy investigations}.

\bibitem[{\citenamefont{Anderson}(1980)}]{anderson1980spinglass6}
\bibinfo{author}{\bibfnamefont{P.~W.} \bibnamefont{Anderson}},
  \bibinfo{journal}{Physics Today} \textbf{\bibinfo{volume}{33}},
  \bibinfo{pages}{9} (\bibinfo{year}{1980}).

\bibitem[{\citenamefont{Aho et~al.}(1974)\citenamefont{Aho, Hopcroft, and
  Ullman}}]{Aho1974}
\bibinfo{author}{\bibfnamefont{A.~V.} \bibnamefont{Aho}},
  \bibinfo{author}{\bibfnamefont{J.~E.} \bibnamefont{Hopcroft}},
  \bibnamefont{and} \bibinfo{author}{\bibfnamefont{J.~D.}
  \bibnamefont{Ullman}}, \emph{\bibinfo{title}{The Design and Analysis of
  Computer Algorithms}} (\bibinfo{publisher}{Addison-Wesley},
  \bibinfo{address}{Reading, Massachusetts}, \bibinfo{year}{1974}).

\bibitem[{\citenamefont{Monasson and Zecchina}(1997)}]{PhysRevE.56.1357}
\bibinfo{author}{\bibfnamefont{R.}~\bibnamefont{Monasson}} \bibnamefont{and}
  \bibinfo{author}{\bibfnamefont{R.}~\bibnamefont{Zecchina}},
  \bibinfo{journal}{Phys. Rev. E} \textbf{\bibinfo{volume}{56}},
  \bibinfo{pages}{1357} (\bibinfo{year}{1997}),
  \urlprefix\url{https://link.aps.org/doi/10.1103/PhysRevE.56.1357}.

\bibitem[{\citenamefont{Hopfield and Tank}(1985)}]{Hopfield1985}
\bibinfo{author}{\bibfnamefont{J.~J.} \bibnamefont{Hopfield}} \bibnamefont{and}
  \bibinfo{author}{\bibfnamefont{D.~W.} \bibnamefont{Tank}},
  \bibinfo{journal}{Biological Cybernetics} \textbf{\bibinfo{volume}{52}},
  \bibinfo{pages}{141} (\bibinfo{year}{1985}),
  \urlprefix\url{https://doi.org/10.1007/BF00339943}.

\bibitem[{\citenamefont{Mohseni et~al.}(2022)\citenamefont{Mohseni, McMahon,
  and Byrnes}}]{Mohseni2022}
\bibinfo{author}{\bibfnamefont{N.}~\bibnamefont{Mohseni}},
  \bibinfo{author}{\bibfnamefont{P.~L.} \bibnamefont{McMahon}},
  \bibnamefont{and} \bibinfo{author}{\bibfnamefont{T.}~\bibnamefont{Byrnes}},
  \bibinfo{journal}{Nature Reviews Physics} \textbf{\bibinfo{volume}{4}},
  \bibinfo{pages}{363} (\bibinfo{year}{2022}),
  \urlprefix\url{https://doi.org/10.1038/s42254-022-00440-8}.

\bibitem[{\citenamefont{Ercsey-Ravasz and Toroczkai}(2011)}]{ErcseyRavasz2011}
\bibinfo{author}{\bibfnamefont{M.}~\bibnamefont{Ercsey-Ravasz}}
  \bibnamefont{and}
  \bibinfo{author}{\bibfnamefont{Z.}~\bibnamefont{Toroczkai}},
  \bibinfo{journal}{Nature Physics} \textbf{\bibinfo{volume}{7}},
  \bibinfo{pages}{966} (\bibinfo{year}{2011}),
  \urlprefix\url{https://doi.org/10.1038/nphys2105}.

\bibitem[{\citenamefont{Wang and Roychowdhury}(2019)}]{inbook}
\bibinfo{author}{\bibfnamefont{T.}~\bibnamefont{Wang}} \bibnamefont{and}
  \bibinfo{author}{\bibfnamefont{J.}~\bibnamefont{Roychowdhury}},
  \emph{\bibinfo{title}{OIM: Oscillator-Based Ising Machines for Solving
  Combinatorial Optimisation Problems}} (\bibinfo{year}{2019}), pp.
  \bibinfo{pages}{232--256}, ISBN \bibinfo{isbn}{978-3-030-19310-2}.

\bibitem[{\citenamefont{Byrnes et~al.}(2013)\citenamefont{Byrnes, Koyama, Yan,
  and Yamamoto}}]{Byrnes2013}
\bibinfo{author}{\bibfnamefont{T.}~\bibnamefont{Byrnes}},
  \bibinfo{author}{\bibfnamefont{S.}~\bibnamefont{Koyama}},
  \bibinfo{author}{\bibfnamefont{K.}~\bibnamefont{Yan}}, \bibnamefont{and}
  \bibinfo{author}{\bibfnamefont{Y.}~\bibnamefont{Yamamoto}},
  \bibinfo{journal}{Scientific Reports} \textbf{\bibinfo{volume}{3}},
  \bibinfo{pages}{2531} (\bibinfo{year}{2013}), ISSN \bibinfo{issn}{2045-2322},
  \urlprefix\url{https://doi.org/10.1038/srep02531}.

\bibitem[{\citenamefont{Chen et~al.}(2025)\citenamefont{Chen, Xiao, Akl,
  Leugering, Olajide, Malik, Dennler, Harper, Bose, Gonzalez
  et~al.}}]{Chen2025}
\bibinfo{author}{\bibfnamefont{Z.}~\bibnamefont{Chen}},
  \bibinfo{author}{\bibfnamefont{Z.}~\bibnamefont{Xiao}},
  \bibinfo{author}{\bibfnamefont{M.}~\bibnamefont{Akl}},
  \bibinfo{author}{\bibfnamefont{J.}~\bibnamefont{Leugering}},
  \bibinfo{author}{\bibfnamefont{O.}~\bibnamefont{Olajide}},
  \bibinfo{author}{\bibfnamefont{A.}~\bibnamefont{Malik}},
  \bibinfo{author}{\bibfnamefont{N.}~\bibnamefont{Dennler}},
  \bibinfo{author}{\bibfnamefont{C.}~\bibnamefont{Harper}},
  \bibinfo{author}{\bibfnamefont{S.}~\bibnamefont{Bose}},
  \bibinfo{author}{\bibfnamefont{H.~A.} \bibnamefont{Gonzalez}},
  \bibnamefont{et~al.}, \bibinfo{journal}{Nature Communications}
  \textbf{\bibinfo{volume}{16}}, \bibinfo{pages}{3086} (\bibinfo{year}{2025}),
  ISSN \bibinfo{issn}{2041-1723},
  \urlprefix\url{https://doi.org/10.1038/s41467-025-58231-5}.

\bibitem[{\citenamefont{Kim et~al.}(2025)\citenamefont{Kim, Son, An, and
  Shim}}]{https://doi.org/10.1049/ell2.70236}
\bibinfo{author}{\bibfnamefont{H.}~\bibnamefont{Kim}},
  \bibinfo{author}{\bibfnamefont{D.}~\bibnamefont{Son}},
  \bibinfo{author}{\bibfnamefont{Y.}~\bibnamefont{An}}, \bibnamefont{and}
  \bibinfo{author}{\bibfnamefont{Y.}~\bibnamefont{Shim}},
  \bibinfo{journal}{Electronics Letters} \textbf{\bibinfo{volume}{61}},
  \bibinfo{pages}{e70236} (\bibinfo{year}{2025}),
  \urlprefix\url{https://ietresearch.onlinelibrary.wiley.com/doi/abs/10.1049/ell2.70236}.

\bibitem[{\citenamefont{Cen et~al.}(2022)\citenamefont{Cen, Ding, Hao, Guan,
  Qin, Lyu, Li, Zhu, Xu, Dai et~al.}}]{Cen2022}
\bibinfo{author}{\bibfnamefont{Q.}~\bibnamefont{Cen}},
  \bibinfo{author}{\bibfnamefont{H.}~\bibnamefont{Ding}},
  \bibinfo{author}{\bibfnamefont{T.}~\bibnamefont{Hao}},
  \bibinfo{author}{\bibfnamefont{S.}~\bibnamefont{Guan}},
  \bibinfo{author}{\bibfnamefont{Z.}~\bibnamefont{Qin}},
  \bibinfo{author}{\bibfnamefont{J.}~\bibnamefont{Lyu}},
  \bibinfo{author}{\bibfnamefont{W.}~\bibnamefont{Li}},
  \bibinfo{author}{\bibfnamefont{N.}~\bibnamefont{Zhu}},
  \bibinfo{author}{\bibfnamefont{K.}~\bibnamefont{Xu}},
  \bibinfo{author}{\bibfnamefont{Y.}~\bibnamefont{Dai}}, \bibnamefont{et~al.},
  \bibinfo{journal}{Light: Science \& Applications}
  \textbf{\bibinfo{volume}{11}}, \bibinfo{pages}{333} (\bibinfo{year}{2022}),
  ISSN \bibinfo{issn}{2047-7538},
  \urlprefix\url{https://doi.org/10.1038/s41377-022-01013-1}.

\bibitem[{\citenamefont{Aadit et~al.}(2022)\citenamefont{Aadit, Grimaldi,
  Carpentieri, Theogarajan, Martinis, Finocchio, and Camsari}}]{Aadit2022}
\bibinfo{author}{\bibfnamefont{N.~A.} \bibnamefont{Aadit}},
  \bibinfo{author}{\bibfnamefont{A.}~\bibnamefont{Grimaldi}},
  \bibinfo{author}{\bibfnamefont{M.}~\bibnamefont{Carpentieri}},
  \bibinfo{author}{\bibfnamefont{L.}~\bibnamefont{Theogarajan}},
  \bibinfo{author}{\bibfnamefont{J.~M.} \bibnamefont{Martinis}},
  \bibinfo{author}{\bibfnamefont{G.}~\bibnamefont{Finocchio}},
  \bibnamefont{and} \bibinfo{author}{\bibfnamefont{K.~Y.}
  \bibnamefont{Camsari}}, \bibinfo{journal}{Nature Electronics}
  \textbf{\bibinfo{volume}{5}}, \bibinfo{pages}{460} (\bibinfo{year}{2022}),
  ISSN \bibinfo{issn}{2520-1131},
  \urlprefix\url{https://doi.org/10.1038/s41928-022-00774-2}.

\bibitem[{\citenamefont{Bybee et~al.}(2023)\citenamefont{Bybee, Kleyko,
  Nikonov, Khosrowshahi, Olshausen, and Sommer}}]{Bybee2023}
\bibinfo{author}{\bibfnamefont{C.}~\bibnamefont{Bybee}},
  \bibinfo{author}{\bibfnamefont{D.}~\bibnamefont{Kleyko}},
  \bibinfo{author}{\bibfnamefont{D.~E.} \bibnamefont{Nikonov}},
  \bibinfo{author}{\bibfnamefont{A.}~\bibnamefont{Khosrowshahi}},
  \bibinfo{author}{\bibfnamefont{B.~A.} \bibnamefont{Olshausen}},
  \bibnamefont{and} \bibinfo{author}{\bibfnamefont{F.~T.}
  \bibnamefont{Sommer}}, \bibinfo{journal}{Nature Communications}
  \textbf{\bibinfo{volume}{14}}, \bibinfo{pages}{6033} (\bibinfo{year}{2023}),
  ISSN \bibinfo{issn}{2041-1723},
  \urlprefix\url{https://doi.org/10.1038/s41467-023-41214-9}.

\bibitem[{\citenamefont{Sejnowski}(1986)}]{10.1063/1.36246}
\bibinfo{author}{\bibfnamefont{T.~J.} \bibnamefont{Sejnowski}},
  \bibinfo{journal}{AIP Conference Proceedings} \textbf{\bibinfo{volume}{151}},
  \bibinfo{pages}{398} (\bibinfo{year}{1986}), ISSN \bibinfo{issn}{0094-243X},
  \eprint{https://pubs.aip.org/aip/acp/article-pdf/151/1/398/12091896/398\_1\_online.pdf},
  \urlprefix\url{https://doi.org/10.1063/1.36246}.

\bibitem[{\citenamefont{{D-Wave Systems Inc.}}(2023)}]{dwave2023guide}
\bibinfo{author}{\bibnamefont{{D-Wave Systems Inc.}}},
  \emph{\bibinfo{title}{Ocean Documentation: The D-Wave Quantum Annealing
  Programming Guide}} (\bibinfo{year}{2023}), \bibinfo{note}{accessed: May
  2025},
  \urlprefix\url{https://www.dwavequantum.com/media/s10ohrq5/dwavedoc_annealing_guide.pdf}.

\bibitem[{\citenamefont{Misra-Spieldenner
  et~al.}(2023)\citenamefont{Misra-Spieldenner, Bode, Schuhmacher, Stollenwerk,
  Bagrets, and Wilhelm}}]{PRXQuantum.4.030335}
\bibinfo{author}{\bibfnamefont{A.}~\bibnamefont{Misra-Spieldenner}},
  \bibinfo{author}{\bibfnamefont{T.}~\bibnamefont{Bode}},
  \bibinfo{author}{\bibfnamefont{P.~K.} \bibnamefont{Schuhmacher}},
  \bibinfo{author}{\bibfnamefont{T.}~\bibnamefont{Stollenwerk}},
  \bibinfo{author}{\bibfnamefont{D.}~\bibnamefont{Bagrets}}, \bibnamefont{and}
  \bibinfo{author}{\bibfnamefont{F.~K.} \bibnamefont{Wilhelm}},
  \bibinfo{journal}{PRX Quantum} \textbf{\bibinfo{volume}{4}},
  \bibinfo{pages}{030335} (\bibinfo{year}{2023}),
  \urlprefix\url{https://link.aps.org/doi/10.1103/PRXQuantum.4.030335}.

\bibitem[{\citenamefont{Jiang et~al.}(2018)\citenamefont{Jiang, Britt,
  McCaskey, Humble, and Kais}}]{Jiang2018}
\bibinfo{author}{\bibfnamefont{S.}~\bibnamefont{Jiang}},
  \bibinfo{author}{\bibfnamefont{K.~A.} \bibnamefont{Britt}},
  \bibinfo{author}{\bibfnamefont{A.~J.} \bibnamefont{McCaskey}},
  \bibinfo{author}{\bibfnamefont{T.~S.} \bibnamefont{Humble}},
  \bibnamefont{and} \bibinfo{author}{\bibfnamefont{S.}~\bibnamefont{Kais}},
  \bibinfo{journal}{Scientific Reports} \textbf{\bibinfo{volume}{8}},
  \bibinfo{pages}{17667} (\bibinfo{year}{2018}), ISSN
  \bibinfo{issn}{2045-2322},
  \urlprefix\url{https://doi.org/10.1038/s41598-018-36058-z}.

\bibitem[{\citenamefont{Wang et~al.}(2020)\citenamefont{Wang, Hu, Yao, and
  Wang}}]{Wang2020}
\bibinfo{author}{\bibfnamefont{B.}~\bibnamefont{Wang}},
  \bibinfo{author}{\bibfnamefont{F.}~\bibnamefont{Hu}},
  \bibinfo{author}{\bibfnamefont{H.}~\bibnamefont{Yao}}, \bibnamefont{and}
  \bibinfo{author}{\bibfnamefont{C.}~\bibnamefont{Wang}},
  \bibinfo{journal}{Scientific Reports} \textbf{\bibinfo{volume}{10}},
  \bibinfo{pages}{7106} (\bibinfo{year}{2020}).

\bibitem[{\citenamefont{Chukwu et~al.}(2020)\citenamefont{Chukwu, Dridi,
  Berwald, Booth, Dawson, Le, Wainger, and Reinhardt}}]{9286230}
\bibinfo{author}{\bibfnamefont{U.}~\bibnamefont{Chukwu}},
  \bibinfo{author}{\bibfnamefont{R.}~\bibnamefont{Dridi}},
  \bibinfo{author}{\bibfnamefont{J.}~\bibnamefont{Berwald}},
  \bibinfo{author}{\bibfnamefont{M.}~\bibnamefont{Booth}},
  \bibinfo{author}{\bibfnamefont{J.}~\bibnamefont{Dawson}},
  \bibinfo{author}{\bibfnamefont{D.}~\bibnamefont{Le}},
  \bibinfo{author}{\bibfnamefont{M.}~\bibnamefont{Wainger}}, \bibnamefont{and}
  \bibinfo{author}{\bibfnamefont{S.~P.} \bibnamefont{Reinhardt}}, in
  \emph{\bibinfo{booktitle}{2020 IEEE High Performance Extreme Computing
  Conference (HPEC)}} (\bibinfo{year}{2020}), pp. \bibinfo{pages}{1--6}.

\bibitem[{\citenamefont{Lin et~al.}(2017)\citenamefont{Lin, Tegmark, and
  Rolnick}}]{Lin2017}
\bibinfo{author}{\bibfnamefont{H.~W.} \bibnamefont{Lin}},
  \bibinfo{author}{\bibfnamefont{M.}~\bibnamefont{Tegmark}}, \bibnamefont{and}
  \bibinfo{author}{\bibfnamefont{D.}~\bibnamefont{Rolnick}},
  \bibinfo{journal}{Journal of Statistical Physics}
  \textbf{\bibinfo{volume}{168}}, \bibinfo{pages}{1223} (\bibinfo{year}{2017}),
  ISSN \bibinfo{issn}{1572-9613},
  \urlprefix\url{https://doi.org/10.1007/s10955-017-1836-5}.

\bibitem[{\citenamefont{Rosenberg}(1972)}]{rosenberg1972breves}
\bibinfo{author}{\bibfnamefont{I.~G.} \bibnamefont{Rosenberg}},
  \bibinfo{journal}{Revue fran{\c{c}}aise d'automatique, informatique,
  recherche op{\'e}rationnelle. Recherche op{\'e rationnelle}}
  \textbf{\bibinfo{volume}{6}}, \bibinfo{pages}{95} (\bibinfo{year}{1972}),
  \bibinfo{note}{accessed: 2025-05-12},
  \urlprefix\url{https://www.numdam.org/item/RO_1972__6_2_95_0/}.

\bibitem[{\citenamefont{Biamonte}(2008)}]{PhysRevA.77.052331}
\bibinfo{author}{\bibfnamefont{J.~D.} \bibnamefont{Biamonte}},
  \bibinfo{journal}{Phys. Rev. A} \textbf{\bibinfo{volume}{77}},
  \bibinfo{pages}{052331} (\bibinfo{year}{2008}),
  \urlprefix\url{https://link.aps.org/doi/10.1103/PhysRevA.77.052331}.

\bibitem[{\citenamefont{Dattani}(2019)}]{dattani2019quadratizationdiscreteoptimizationquantum}
\bibinfo{author}{\bibfnamefont{N.}~\bibnamefont{Dattani}},
  \emph{\bibinfo{title}{Quadratization in discrete optimization and quantum
  mechanics}} (\bibinfo{year}{2019}), \eprint{1901.04405},
  \urlprefix\url{https://arxiv.org/abs/1901.04405}.

\bibitem[{\citenamefont{Anthony et~al.}(2017)\citenamefont{Anthony, Boros,
  Crama, and Gruber}}]{Anthony2017}
\bibinfo{author}{\bibfnamefont{M.}~\bibnamefont{Anthony}},
  \bibinfo{author}{\bibfnamefont{E.}~\bibnamefont{Boros}},
  \bibinfo{author}{\bibfnamefont{Y.}~\bibnamefont{Crama}}, \bibnamefont{and}
  \bibinfo{author}{\bibfnamefont{A.}~\bibnamefont{Gruber}},
  \bibinfo{journal}{Mathematical Programming} \textbf{\bibinfo{volume}{162}},
  \bibinfo{pages}{115} (\bibinfo{year}{2017}), ISSN \bibinfo{issn}{1436-4646},
  \urlprefix\url{https://doi.org/10.1007/s10107-016-1032-4}.

\bibitem[{\citenamefont{Dobrynin
  et~al.}(2024{\natexlab{a}})\citenamefont{Dobrynin, Renaudineau, Hizzani,
  Strukov, Mohseni, and Strachan}}]{PhysRevE.110.045308}
\bibinfo{author}{\bibfnamefont{D.}~\bibnamefont{Dobrynin}},
  \bibinfo{author}{\bibfnamefont{A.}~\bibnamefont{Renaudineau}},
  \bibinfo{author}{\bibfnamefont{M.}~\bibnamefont{Hizzani}},
  \bibinfo{author}{\bibfnamefont{D.}~\bibnamefont{Strukov}},
  \bibinfo{author}{\bibfnamefont{M.}~\bibnamefont{Mohseni}}, \bibnamefont{and}
  \bibinfo{author}{\bibfnamefont{J.~P.} \bibnamefont{Strachan}},
  \bibinfo{journal}{Phys. Rev. E} \textbf{\bibinfo{volume}{110}},
  \bibinfo{pages}{045308} (\bibinfo{year}{2024}{\natexlab{a}}),
  \urlprefix\url{https://link.aps.org/doi/10.1103/PhysRevE.110.045308}.

\bibitem[{\citenamefont{Hizzani et~al.}(2024)\citenamefont{Hizzani, Heittmann,
  Hutchinson, Dobrynin, Van~Vaerenbergh, Bhattacharya, Renaudineau, Strukov,
  and Strachan}}]{10558658}
\bibinfo{author}{\bibfnamefont{M.}~\bibnamefont{Hizzani}},
  \bibinfo{author}{\bibfnamefont{A.}~\bibnamefont{Heittmann}},
  \bibinfo{author}{\bibfnamefont{G.}~\bibnamefont{Hutchinson}},
  \bibinfo{author}{\bibfnamefont{D.}~\bibnamefont{Dobrynin}},
  \bibinfo{author}{\bibfnamefont{T.}~\bibnamefont{Van~Vaerenbergh}},
  \bibinfo{author}{\bibfnamefont{T.}~\bibnamefont{Bhattacharya}},
  \bibinfo{author}{\bibfnamefont{A.}~\bibnamefont{Renaudineau}},
  \bibinfo{author}{\bibfnamefont{D.}~\bibnamefont{Strukov}}, \bibnamefont{and}
  \bibinfo{author}{\bibfnamefont{J.~P.} \bibnamefont{Strachan}}, in
  \emph{\bibinfo{booktitle}{2024 IEEE International Symposium on Circuits and
  Systems (ISCAS)}} (\bibinfo{year}{2024}), pp. \bibinfo{pages}{1--5}.

\bibitem[{\citenamefont{Kauffman and Levin}(1987)}]{KAUFFMAN198711}
\bibinfo{author}{\bibfnamefont{S.}~\bibnamefont{Kauffman}} \bibnamefont{and}
  \bibinfo{author}{\bibfnamefont{S.}~\bibnamefont{Levin}},
  \bibinfo{journal}{Journal of Theoretical Biology}
  \textbf{\bibinfo{volume}{128}}, \bibinfo{pages}{11} (\bibinfo{year}{1987}),
  ISSN \bibinfo{issn}{0022-5193},
  \urlprefix\url{https://www.sciencedirect.com/science/article/pii/S0022519387800292}.

\bibitem[{\citenamefont{Sherrington and
  Kirkpatrick}(1975)}]{PhysRevLett.35.1792}
\bibinfo{author}{\bibfnamefont{D.}~\bibnamefont{Sherrington}} \bibnamefont{and}
  \bibinfo{author}{\bibfnamefont{S.}~\bibnamefont{Kirkpatrick}},
  \bibinfo{journal}{Phys. Rev. Lett.} \textbf{\bibinfo{volume}{35}},
  \bibinfo{pages}{1792} (\bibinfo{year}{1975}),
  \urlprefix\url{https://link.aps.org/doi/10.1103/PhysRevLett.35.1792}.

\bibitem[{\citenamefont{Foos et~al.}(2025)\citenamefont{Foos, Epping, Grundler,
  Ciobanu, Singh, Bode, Helias, and Dahmen}}]{foos2025beyond}
\bibinfo{author}{\bibfnamefont{N.}~\bibnamefont{Foos}},
  \bibinfo{author}{\bibfnamefont{B.}~\bibnamefont{Epping}},
  \bibinfo{author}{\bibfnamefont{J.}~\bibnamefont{Grundler}},
  \bibinfo{author}{\bibfnamefont{A.}~\bibnamefont{Ciobanu}},
  \bibinfo{author}{\bibfnamefont{A.}~\bibnamefont{Singh}},
  \bibinfo{author}{\bibfnamefont{T.}~\bibnamefont{Bode}},
  \bibinfo{author}{\bibfnamefont{M.}~\bibnamefont{Helias}}, \bibnamefont{and}
  \bibinfo{author}{\bibfnamefont{D.}~\bibnamefont{Dahmen}}
  (\bibinfo{year}{2025}), \bibinfo{note}{arXiv:2507.10360 [cond-mat.dis-nn]},
  \urlprefix\url{https://arxiv.org/abs/2507.10360}.

\bibitem[{\citenamefont{Pinkas}(1991)}]{10.1162/neco.1991.3.2.282}
\bibinfo{author}{\bibfnamefont{G.}~\bibnamefont{Pinkas}},
  \bibinfo{journal}{Neural Computation} \textbf{\bibinfo{volume}{3}},
  \bibinfo{pages}{282} (\bibinfo{year}{1991}), ISSN \bibinfo{issn}{0899-7667},
  \eprint{https://direct.mit.edu/neco/article-pdf/3/2/282/812143/neco.1991.3.2.282.pdf},
  \urlprefix\url{https://doi.org/10.1162/neco.1991.3.2.282}.

\bibitem[{\citenamefont{van Saarloos}(1983)}]{PhysRevB.27.5678}
\bibinfo{author}{\bibfnamefont{W.}~\bibnamefont{van Saarloos}},
  \bibinfo{journal}{Phys. Rev. B} \textbf{\bibinfo{volume}{27}},
  \bibinfo{pages}{5678} (\bibinfo{year}{1983}),
  \urlprefix\url{https://link.aps.org/doi/10.1103/PhysRevB.27.5678}.

\bibitem[{\citenamefont{Walker}(2023)}]{Walker_2023}
\bibinfo{author}{\bibfnamefont{J.~S.} \bibnamefont{Walker}},
  \emph{\bibinfo{title}{Position-Space Renormalization-Group Techniques}}
  (\bibinfo{publisher}{Cambridge University Press}, \bibinfo{year}{2023}), pp.
  \bibinfo{pages}{86--88}, Student{\textquotesingle}s Guides.

\bibitem[{\citenamefont{Wilson}(1971)}]{PhysRevB.4.3174}
\bibinfo{author}{\bibfnamefont{K.~G.} \bibnamefont{Wilson}},
  \bibinfo{journal}{Phys. Rev. B} \textbf{\bibinfo{volume}{4}},
  \bibinfo{pages}{3174} (\bibinfo{year}{1971}),
  \urlprefix\url{https://link.aps.org/doi/10.1103/PhysRevB.4.3174}.

\bibitem[{\citenamefont{Kadanoff}(1976)}]{KADANOFF1976359}
\bibinfo{author}{\bibfnamefont{L.~P.} \bibnamefont{Kadanoff}},
  \bibinfo{journal}{Annals of Physics} \textbf{\bibinfo{volume}{100}},
  \bibinfo{pages}{359} (\bibinfo{year}{1976}), ISSN \bibinfo{issn}{0003-4916},
  \urlprefix\url{https://www.sciencedirect.com/science/article/pii/000349167690066X}.

\bibitem[{\citenamefont{Goldstein and Walker}(1983)}]{10.1063/1.444839}
\bibinfo{author}{\bibfnamefont{R.~E.} \bibnamefont{Goldstein}}
  \bibnamefont{and} \bibinfo{author}{\bibfnamefont{J.~S.}
  \bibnamefont{Walker}}, \bibinfo{journal}{The Journal of Chemical Physics}
  \textbf{\bibinfo{volume}{78}}, \bibinfo{pages}{1492} (\bibinfo{year}{1983}),
  ISSN \bibinfo{issn}{0021-9606},
  \eprint{https://pubs.aip.org/aip/jcp/article-pdf/78/3/1492/18940883/1492\_1\_online.pdf},
  \urlprefix\url{https://doi.org/10.1063/1.444839}.

\bibitem[{\citenamefont{Fisher}(1998)}]{RevModPhys.70.653}
\bibinfo{author}{\bibfnamefont{M.~E.} \bibnamefont{Fisher}},
  \bibinfo{journal}{Rev. Mod. Phys.} \textbf{\bibinfo{volume}{70}},
  \bibinfo{pages}{653} (\bibinfo{year}{1998}),
  \urlprefix\url{https://link.aps.org/doi/10.1103/RevModPhys.70.653}.

\bibitem[{\citenamefont{van Leeuwen}(1978)}]{10.1007/3-540-08848-2_17}
\bibinfo{author}{\bibfnamefont{J.~M.~J.} \bibnamefont{van Leeuwen}}, in
  \emph{\bibinfo{booktitle}{Group Theoretical Methods in Physics}}, edited by
  \bibinfo{editor}{\bibfnamefont{P.}~\bibnamefont{Kramer}} \bibnamefont{and}
  \bibinfo{editor}{\bibfnamefont{A.}~\bibnamefont{Rieckers}}
  (\bibinfo{publisher}{Springer Berlin Heidelberg}, \bibinfo{address}{Berlin,
  Heidelberg}, \bibinfo{year}{1978}), pp. \bibinfo{pages}{325--342}, ISBN
  \bibinfo{isbn}{978-3-540-35813-8}.

\bibitem[{\citenamefont{Beretta et~al.}(2018)\citenamefont{Beretta, Battistin,
  De~Mulatier, Mastromatteo, and Marsili}}]{e20100739}
\bibinfo{author}{\bibfnamefont{A.}~\bibnamefont{Beretta}},
  \bibinfo{author}{\bibfnamefont{C.}~\bibnamefont{Battistin}},
  \bibinfo{author}{\bibfnamefont{C.}~\bibnamefont{De~Mulatier}},
  \bibinfo{author}{\bibfnamefont{I.}~\bibnamefont{Mastromatteo}},
  \bibnamefont{and} \bibinfo{author}{\bibfnamefont{M.}~\bibnamefont{Marsili}},
  \bibinfo{journal}{Entropy} \textbf{\bibinfo{volume}{20}}
  (\bibinfo{year}{2018}), ISSN \bibinfo{issn}{1099-4300},
  \urlprefix\url{https://www.mdpi.com/1099-4300/20/10/739}.

\bibitem[{\citenamefont{las Cuevas and
  Cubitt}(2016)}]{doi:10.1126/science.aab3326}
\bibinfo{author}{\bibfnamefont{G.~D.} \bibnamefont{las Cuevas}}
  \bibnamefont{and} \bibinfo{author}{\bibfnamefont{T.~S.}
  \bibnamefont{Cubitt}}, \bibinfo{journal}{Science}
  \textbf{\bibinfo{volume}{351}}, \bibinfo{pages}{1180} (\bibinfo{year}{2016}),
  \eprint{https://www.science.org/doi/pdf/10.1126/science.aab3326},
  \urlprefix\url{https://www.science.org/doi/abs/10.1126/science.aab3326}.

\bibitem[{\citenamefont{M{\"u}ller}(2010)}]{Mueller2010}
\bibinfo{author}{\bibfnamefont{M.}~\bibnamefont{M{\"u}ller}},
  \emph{\bibinfo{title}{Optimized parallel tempering monte carlo}},
  \bibinfo{howpublished}{Guest Student Programme 2010 presentation, J{\"u}lich
  Supercomputing Centre, Forschungszentrum J{\"u}lich} (\bibinfo{year}{2010}),
  \bibinfo{note}{available at:
  \url{https://www.fz-juelich.de/en/ias/jsc/education/master-students/guest-student-programme/2010-guest-student-programme/11_mueller/@@download/file}
  (accessed 12~August~2025)}.

\bibitem[{\citenamefont{Swendsen and Wang}(1986)}]{PhysRevLett.57.2607}
\bibinfo{author}{\bibfnamefont{R.~H.} \bibnamefont{Swendsen}} \bibnamefont{and}
  \bibinfo{author}{\bibfnamefont{J.-S.} \bibnamefont{Wang}},
  \bibinfo{journal}{Phys. Rev. Lett.} \textbf{\bibinfo{volume}{57}},
  \bibinfo{pages}{2607} (\bibinfo{year}{1986}),
  \urlprefix\url{https://link.aps.org/doi/10.1103/PhysRevLett.57.2607}.

\bibitem[{\citenamefont{Hukushima and Nemoto}(1996)}]{doi:10.1143/JPSJ.65.1604}
\bibinfo{author}{\bibfnamefont{K.}~\bibnamefont{Hukushima}} \bibnamefont{and}
  \bibinfo{author}{\bibfnamefont{K.}~\bibnamefont{Nemoto}},
  \bibinfo{journal}{Journal of the Physical Society of Japan}
  \textbf{\bibinfo{volume}{65}}, \bibinfo{pages}{1604} (\bibinfo{year}{1996}),
  \eprint{https://doi.org/10.1143/JPSJ.65.1604},
  \urlprefix\url{https://doi.org/10.1143/JPSJ.65.1604}.

\bibitem[{\citenamefont{Glauber}(1963)}]{Glauber1963}
\bibinfo{author}{\bibfnamefont{R.~J.} \bibnamefont{Glauber}},
  \bibinfo{journal}{J. Math. Phys.} \textbf{\bibinfo{volume}{4}},
  \bibinfo{pages}{294} (\bibinfo{year}{1963}),
  \urlprefix\url{https://doi.org/10.1063/1.1703954}.

\bibitem[{\citenamefont{M\'{e}zard et~al.}(2002)\citenamefont{M\'{e}zard,
  Parisi, and Zecchina}}]{Mezard2002}
\bibinfo{author}{\bibfnamefont{M.}~\bibnamefont{M\'{e}zard}},
  \bibinfo{author}{\bibfnamefont{G.}~\bibnamefont{Parisi}}, \bibnamefont{and}
  \bibinfo{author}{\bibfnamefont{R.}~\bibnamefont{Zecchina}},
  \bibinfo{journal}{Science} \textbf{\bibinfo{volume}{297}},
  \bibinfo{pages}{812} (\bibinfo{year}{2002}),
  \eprint{https://www.science.org/doi/pdf/10.1126/science.1073287},
  \urlprefix\url{https://www.science.org/doi/abs/10.1126/science.1073287}.

\bibitem[{\citenamefont{Dobrynin
  et~al.}(2024{\natexlab{b}})\citenamefont{Dobrynin, Tedeschi, Heittmann, and
  Strachan}}]{10937013}
\bibinfo{author}{\bibfnamefont{D.}~\bibnamefont{Dobrynin}},
  \bibinfo{author}{\bibfnamefont{M.}~\bibnamefont{Tedeschi}},
  \bibinfo{author}{\bibfnamefont{A.}~\bibnamefont{Heittmann}},
  \bibnamefont{and} \bibinfo{author}{\bibfnamefont{J.~P.}
  \bibnamefont{Strachan}}, in \emph{\bibinfo{booktitle}{2024 IEEE International
  Conference on Rebooting Computing (ICRC)}}
  (\bibinfo{year}{2024}{\natexlab{b}}), pp. \bibinfo{pages}{1--10}.

\bibitem[{\citenamefont{M\'{e}zard and Parisi}(2001)}]{Mezard2001}
\bibinfo{author}{\bibfnamefont{M.}~\bibnamefont{M\'{e}zard}} \bibnamefont{and}
  \bibinfo{author}{\bibfnamefont{G.}~\bibnamefont{Parisi}},
  \bibinfo{journal}{Eur. Phys. J. B} \textbf{\bibinfo{volume}{20}},
  \bibinfo{pages}{217} (\bibinfo{year}{2001}),
  \urlprefix\url{https://doi.org/10.1007/PL00011099}.

\bibitem[{\citenamefont{Kuwahara}(2025)}]{9hx7-pzxw}
\bibinfo{author}{\bibfnamefont{T.}~\bibnamefont{Kuwahara}},
  \bibinfo{journal}{Phys. Rev. X} pp.~\bibinfo{pages}{--}
  (\bibinfo{year}{2025}),
  \urlprefix\url{https://link.aps.org/doi/10.1103/9hx7-pzxw}.

\bibitem[{\citenamefont{Braunstein et~al.}(2005)\citenamefont{Braunstein,
  M{\'e}zard, and Zecchina}}]{Braunstein2005}
\bibinfo{author}{\bibfnamefont{A.}~\bibnamefont{Braunstein}},
  \bibinfo{author}{\bibfnamefont{M.}~\bibnamefont{M{\'e}zard}},
  \bibnamefont{and} \bibinfo{author}{\bibfnamefont{R.}~\bibnamefont{Zecchina}},
  \bibinfo{journal}{Random Structures \& Algorithms}
  \textbf{\bibinfo{volume}{27}}, \bibinfo{pages}{201} (\bibinfo{year}{2005}),
  \eprint{https://onlinelibrary.wiley.com/doi/pdf/10.1002/rsa.20057},
  \urlprefix\url{https://onlinelibrary.wiley.com/doi/abs/10.1002/rsa.20057}.

\bibitem[{\citenamefont{Cai et~al.}(2020)\citenamefont{Cai, Kumar, Vaerenbergh,
  Sheng, Liu, Li, Liu, Foltin, Yu, Xia et~al.}}]{article_cai}
\bibinfo{author}{\bibfnamefont{F.}~\bibnamefont{Cai}},
  \bibinfo{author}{\bibfnamefont{S.}~\bibnamefont{Kumar}},
  \bibinfo{author}{\bibfnamefont{T.}~\bibnamefont{Vaerenbergh}},
  \bibinfo{author}{\bibfnamefont{X.}~\bibnamefont{Sheng}},
  \bibinfo{author}{\bibfnamefont{R.}~\bibnamefont{Liu}},
  \bibinfo{author}{\bibfnamefont{C.}~\bibnamefont{Li}},
  \bibinfo{author}{\bibfnamefont{Z.}~\bibnamefont{Liu}},
  \bibinfo{author}{\bibfnamefont{M.}~\bibnamefont{Foltin}},
  \bibinfo{author}{\bibfnamefont{S.}~\bibnamefont{Yu}},
  \bibinfo{author}{\bibfnamefont{Q.}~\bibnamefont{Xia}}, \bibnamefont{et~al.},
  \bibinfo{journal}{Nature Electronics} \textbf{\bibinfo{volume}{3}},
  \bibinfo{pages}{1} (\bibinfo{year}{2020}).

\bibitem[{\citenamefont{Pedretti et~al.}(2025)\citenamefont{Pedretti, B{\"o}hm,
  Bhattacharya, Heittmann, Zhang, Hizzani, Hutchinson, Kwon, Moon, Valiante
  et~al.}}]{Pedretti2025}
\bibinfo{author}{\bibfnamefont{G.}~\bibnamefont{Pedretti}},
  \bibinfo{author}{\bibfnamefont{F.}~\bibnamefont{B{\"o}hm}},
  \bibinfo{author}{\bibfnamefont{T.}~\bibnamefont{Bhattacharya}},
  \bibinfo{author}{\bibfnamefont{A.}~\bibnamefont{Heittmann}},
  \bibinfo{author}{\bibfnamefont{X.}~\bibnamefont{Zhang}},
  \bibinfo{author}{\bibfnamefont{M.}~\bibnamefont{Hizzani}},
  \bibinfo{author}{\bibfnamefont{G.}~\bibnamefont{Hutchinson}},
  \bibinfo{author}{\bibfnamefont{D.}~\bibnamefont{Kwon}},
  \bibinfo{author}{\bibfnamefont{J.}~\bibnamefont{Moon}},
  \bibinfo{author}{\bibfnamefont{E.}~\bibnamefont{Valiante}},
  \bibnamefont{et~al.}, \bibinfo{journal}{npj Unconventional Computing}
  \textbf{\bibinfo{volume}{2}}, \bibinfo{pages}{7} (\bibinfo{year}{2025}), ISSN
  \bibinfo{issn}{3004-8672},
  \urlprefix\url{https://doi.org/10.1038/s44335-025-00020-w}.

\bibitem[{\citenamefont{Hoos}()}]{hoos_satlib}
\bibinfo{author}{\bibfnamefont{H.~H.} \bibnamefont{Hoos}},
  \emph{\bibinfo{title}{Satlib benchmark problems: Uniform random}},
  \bibinfo{howpublished}{\url{https://www.cs.ubc.ca/~hoos/SATLIB/benchm.html}},
  \bibinfo{note}{accessed: 2025-08-05}.

\bibitem[{\citenamefont{Tsukamoto et~al.}(2017)\citenamefont{Tsukamoto,
  Takatsu, Matsubara, and Tamura}}]{article_fj}
\bibinfo{author}{\bibfnamefont{S.}~\bibnamefont{Tsukamoto}},
  \bibinfo{author}{\bibfnamefont{M.}~\bibnamefont{Takatsu}},
  \bibinfo{author}{\bibfnamefont{S.}~\bibnamefont{Matsubara}},
  \bibnamefont{and} \bibinfo{author}{\bibfnamefont{H.}~\bibnamefont{Tamura}},
  \bibinfo{journal}{Fujitsu Scientific and Technical Journal}
  \textbf{\bibinfo{volume}{53}}, \bibinfo{pages}{8} (\bibinfo{year}{2017}).

\bibitem[{\citenamefont{Sasaki et~al.}(2023)\citenamefont{Sasaki, Nakagawa,
  Mimura, Okada, Aonishi, and Yamamoto}}]{doi:10.7566/JPSJ.92.044002}
\bibinfo{author}{\bibfnamefont{R.}~\bibnamefont{Sasaki}},
  \bibinfo{author}{\bibfnamefont{S.}~\bibnamefont{Nakagawa}},
  \bibinfo{author}{\bibfnamefont{K.}~\bibnamefont{Mimura}},
  \bibinfo{author}{\bibfnamefont{M.}~\bibnamefont{Okada}},
  \bibinfo{author}{\bibfnamefont{T.}~\bibnamefont{Aonishi}}, \bibnamefont{and}
  \bibinfo{author}{\bibfnamefont{Y.}~\bibnamefont{Yamamoto}},
  \bibinfo{journal}{Journal of the Physical Society of Japan}
  \textbf{\bibinfo{volume}{92}}, \bibinfo{pages}{044002}
  (\bibinfo{year}{2023}), \eprint{https://doi.org/10.7566/JPSJ.92.044002},
  \urlprefix\url{https://doi.org/10.7566/JPSJ.92.044002}.

\bibitem[{\citenamefont{Honjo et~al.}(2021)\citenamefont{Honjo, Sonobe, Inaba,
  Inagaki, Ikuta, Yamada, Kazama, Enbutsu, Umeki, Kasahara
  et~al.}}]{doi:10.1126/sciadv.abh0952}
\bibinfo{author}{\bibfnamefont{T.}~\bibnamefont{Honjo}},
  \bibinfo{author}{\bibfnamefont{T.}~\bibnamefont{Sonobe}},
  \bibinfo{author}{\bibfnamefont{K.}~\bibnamefont{Inaba}},
  \bibinfo{author}{\bibfnamefont{T.}~\bibnamefont{Inagaki}},
  \bibinfo{author}{\bibfnamefont{T.}~\bibnamefont{Ikuta}},
  \bibinfo{author}{\bibfnamefont{Y.}~\bibnamefont{Yamada}},
  \bibinfo{author}{\bibfnamefont{T.}~\bibnamefont{Kazama}},
  \bibinfo{author}{\bibfnamefont{K.}~\bibnamefont{Enbutsu}},
  \bibinfo{author}{\bibfnamefont{T.}~\bibnamefont{Umeki}},
  \bibinfo{author}{\bibfnamefont{R.}~\bibnamefont{Kasahara}},
  \bibnamefont{et~al.}, \bibinfo{journal}{Science Advances}
  \textbf{\bibinfo{volume}{7}}, \bibinfo{pages}{eabh0952}
  (\bibinfo{year}{2021}),
  \eprint{https://www.science.org/doi/pdf/10.1126/sciadv.abh0952},
  \urlprefix\url{https://www.science.org/doi/abs/10.1126/sciadv.abh0952}.

\bibitem[{\citenamefont{Goto et~al.}(2021)\citenamefont{Goto, Endo, Suzuki,
  Sakai, Kanao, Hamakawa, Hidaka, Yamasaki, and
  Tatsumura}}]{doi:10.1126/sciadv.abe7953}
\bibinfo{author}{\bibfnamefont{H.}~\bibnamefont{Goto}},
  \bibinfo{author}{\bibfnamefont{K.}~\bibnamefont{Endo}},
  \bibinfo{author}{\bibfnamefont{M.}~\bibnamefont{Suzuki}},
  \bibinfo{author}{\bibfnamefont{Y.}~\bibnamefont{Sakai}},
  \bibinfo{author}{\bibfnamefont{T.}~\bibnamefont{Kanao}},
  \bibinfo{author}{\bibfnamefont{Y.}~\bibnamefont{Hamakawa}},
  \bibinfo{author}{\bibfnamefont{R.}~\bibnamefont{Hidaka}},
  \bibinfo{author}{\bibfnamefont{M.}~\bibnamefont{Yamasaki}}, \bibnamefont{and}
  \bibinfo{author}{\bibfnamefont{K.}~\bibnamefont{Tatsumura}},
  \bibinfo{journal}{Science Advances} \textbf{\bibinfo{volume}{7}},
  \bibinfo{pages}{eabe7953} (\bibinfo{year}{2021}),
  \eprint{https://www.science.org/doi/pdf/10.1126/sciadv.abe7953},
  \urlprefix\url{https://www.science.org/doi/abs/10.1126/sciadv.abe7953}.

\bibitem[{\citenamefont{English et~al.}(2022)\citenamefont{English, Zampetaki,
  Kalinin, Berloff, and Kevrekidis}}]{English2022}
\bibinfo{author}{\bibfnamefont{L.~Q.} \bibnamefont{English}},
  \bibinfo{author}{\bibfnamefont{A.~V.} \bibnamefont{Zampetaki}},
  \bibinfo{author}{\bibfnamefont{K.~P.} \bibnamefont{Kalinin}},
  \bibinfo{author}{\bibfnamefont{N.~G.} \bibnamefont{Berloff}},
  \bibnamefont{and} \bibinfo{author}{\bibfnamefont{P.~G.}
  \bibnamefont{Kevrekidis}}, \bibinfo{journal}{Communications Physics}
  \textbf{\bibinfo{volume}{5}}, \bibinfo{pages}{333} (\bibinfo{year}{2022}),
  ISSN \bibinfo{issn}{2399-3650},
  \urlprefix\url{https://doi.org/10.1038/s42005-022-01111-x}.

\bibitem[{\citenamefont{Jiang et~al.}(2023)\citenamefont{Jiang, Shan, He, and
  Li}}]{Jiang2023}
\bibinfo{author}{\bibfnamefont{M.}~\bibnamefont{Jiang}},
  \bibinfo{author}{\bibfnamefont{K.}~\bibnamefont{Shan}},
  \bibinfo{author}{\bibfnamefont{C.}~\bibnamefont{He}}, \bibnamefont{and}
  \bibinfo{author}{\bibfnamefont{C.}~\bibnamefont{Li}},
  \bibinfo{journal}{Nature Communications} \textbf{\bibinfo{volume}{14}},
  \bibinfo{pages}{5927} (\bibinfo{year}{2023}), ISSN \bibinfo{issn}{2041-1723},
  \urlprefix\url{https://doi.org/10.1038/s41467-023-41647-2}.

\bibitem[{\citenamefont{Kalinin and Berloff}(2018)}]{Kalinin2018}
\bibinfo{author}{\bibfnamefont{K.~P.} \bibnamefont{Kalinin}} \bibnamefont{and}
  \bibinfo{author}{\bibfnamefont{N.~G.} \bibnamefont{Berloff}},
  \bibinfo{journal}{Scientific Reports} \textbf{\bibinfo{volume}{8}},
  \bibinfo{pages}{17791} (\bibinfo{year}{2018}), ISSN
  \bibinfo{issn}{2045-2322},
  \urlprefix\url{https://doi.org/10.1038/s41598-018-35416-1}.

\bibitem[{\citenamefont{Prins et~al.}(2025)\citenamefont{Prins, der Sande,
  Bienstman, and
  Vaerenbergh}}]{deprins2025incorporatehigherorderinteractionsanalog}
\bibinfo{author}{\bibfnamefont{R.~D.} \bibnamefont{Prins}},
  \bibinfo{author}{\bibfnamefont{G.~V.} \bibnamefont{der Sande}},
  \bibinfo{author}{\bibfnamefont{P.}~\bibnamefont{Bienstman}},
  \bibnamefont{and} \bibinfo{author}{\bibfnamefont{T.~V.}
  \bibnamefont{Vaerenbergh}}, \emph{\bibinfo{title}{How to incorporate
  higher-order interactions in analog ising machines}} (\bibinfo{year}{2025}),
  \eprint{2507.23621}, \urlprefix\url{https://arxiv.org/abs/2507.23621}.

\bibitem[{\citenamefont{Drieb-Sch{\"{o}}n
  et~al.}(2023)\citenamefont{Drieb-Sch{\"{o}}n, Ender, Javanmard, and
  Lechner}}]{DriebSchon2023parityquantum}
\bibinfo{author}{\bibfnamefont{M.}~\bibnamefont{Drieb-Sch{\"{o}}n}},
  \bibinfo{author}{\bibfnamefont{K.}~\bibnamefont{Ender}},
  \bibinfo{author}{\bibfnamefont{Y.}~\bibnamefont{Javanmard}},
  \bibnamefont{and} \bibinfo{author}{\bibfnamefont{W.}~\bibnamefont{Lechner}},
  \bibinfo{journal}{{Quantum}} \textbf{\bibinfo{volume}{7}},
  \bibinfo{pages}{951} (\bibinfo{year}{2023}), ISSN \bibinfo{issn}{2521-327X},
  \urlprefix\url{https://doi.org/10.22331/q-2023-03-17-951}.

\bibitem[{\citenamefont{Lanthaler et~al.}(2023)\citenamefont{Lanthaler, Dlaska,
  Ender, and Lechner}}]{PhysRevLett.130.220601}
\bibinfo{author}{\bibfnamefont{M.}~\bibnamefont{Lanthaler}},
  \bibinfo{author}{\bibfnamefont{C.}~\bibnamefont{Dlaska}},
  \bibinfo{author}{\bibfnamefont{K.}~\bibnamefont{Ender}}, \bibnamefont{and}
  \bibinfo{author}{\bibfnamefont{W.}~\bibnamefont{Lechner}},
  \bibinfo{journal}{Phys. Rev. Lett.} \textbf{\bibinfo{volume}{130}},
  \bibinfo{pages}{220601} (\bibinfo{year}{2023}),
  \urlprefix\url{https://link.aps.org/doi/10.1103/PhysRevLett.130.220601}.

\bibitem[{\citenamefont{Chang and Guo}(2003)}]{CHANG2003263}
\bibinfo{author}{\bibfnamefont{W.-L.} \bibnamefont{Chang}} \bibnamefont{and}
  \bibinfo{author}{\bibfnamefont{M.}~\bibnamefont{Guo}},
  \bibinfo{journal}{Biosystems} \textbf{\bibinfo{volume}{72}},
  \bibinfo{pages}{263} (\bibinfo{year}{2003}), ISSN \bibinfo{issn}{0303-2647},
  \urlprefix\url{https://www.sciencedirect.com/science/article/pii/S0303264703001497}.

\bibitem[{\citenamefont{Cygan}(2013)}]{article}
\bibinfo{author}{\bibfnamefont{M.}~\bibnamefont{Cygan}},
  \bibinfo{journal}{Proceedings - Annual IEEE Symposium on Foundations of
  Computer Science, FOCS}  (\bibinfo{year}{2013}).

\bibitem[{\citenamefont{Cao et~al.}(2016)\citenamefont{Cao, Jiang, Perouli, and
  Kais}}]{Cao2016}
\bibinfo{author}{\bibfnamefont{Y.}~\bibnamefont{Cao}},
  \bibinfo{author}{\bibfnamefont{S.}~\bibnamefont{Jiang}},
  \bibinfo{author}{\bibfnamefont{D.}~\bibnamefont{Perouli}}, \bibnamefont{and}
  \bibinfo{author}{\bibfnamefont{S.}~\bibnamefont{Kais}},
  \bibinfo{journal}{Scientific Reports} \textbf{\bibinfo{volume}{6}},
  \bibinfo{pages}{33957} (\bibinfo{year}{2016}),
  \urlprefix\url{https://doi.org/10.1038/srep33957}.

\bibitem[{\citenamefont{Singh et~al.}(2025)\citenamefont{Singh, Chandrasekar,
  Zou, Kurths, and Senthilkumar}}]{Singh2025}
\bibinfo{author}{\bibfnamefont{K.}~\bibnamefont{Singh}},
  \bibinfo{author}{\bibfnamefont{V.~K.} \bibnamefont{Chandrasekar}},
  \bibinfo{author}{\bibfnamefont{W.}~\bibnamefont{Zou}},
  \bibinfo{author}{\bibfnamefont{J.}~\bibnamefont{Kurths}}, \bibnamefont{and}
  \bibinfo{author}{\bibfnamefont{D.~V.} \bibnamefont{Senthilkumar}},
  \bibinfo{journal}{Communications Physics} \textbf{\bibinfo{volume}{8}},
  \bibinfo{pages}{170} (\bibinfo{year}{2025}),
  \urlprefix\url{https://doi.org/10.1038/s42005-025-02089-y}.

\bibitem[{\citenamefont{Singh et~al.}(2024)\citenamefont{Singh, Lloyd, and
  Flicker}}]{PhysRevX.14.031005}
\bibinfo{author}{\bibfnamefont{S.}~\bibnamefont{Singh}},
  \bibinfo{author}{\bibfnamefont{J.}~\bibnamefont{Lloyd}}, \bibnamefont{and}
  \bibinfo{author}{\bibfnamefont{F.}~\bibnamefont{Flicker}},
  \bibinfo{journal}{Phys. Rev. X} \textbf{\bibinfo{volume}{14}},
  \bibinfo{pages}{031005} (\bibinfo{year}{2024}),
  \urlprefix\url{https://link.aps.org/doi/10.1103/PhysRevX.14.031005}.

\bibitem[{\citenamefont{Karimzadehgan and Zhai}(2012)}]{Karimzadehgan2012}
\bibinfo{author}{\bibfnamefont{M.}~\bibnamefont{Karimzadehgan}}
  \bibnamefont{and} \bibinfo{author}{\bibfnamefont{C.}~\bibnamefont{Zhai}},
  \bibinfo{journal}{Information Processing \& Management}
  \textbf{\bibinfo{volume}{48}}, \bibinfo{pages}{725} (\bibinfo{year}{2012}),
  \urlprefix\url{https://doi.org/10.1016/j.ipm.2011.09.004}.

\bibitem[{\citenamefont{J\"{u}lich and Centre}(2021)}]{Julich2021}
\bibinfo{author}{\bibfnamefont{F.}~\bibnamefont{J\"{u}lich}} \bibnamefont{and}
  \bibinfo{author}{\bibfnamefont{J.~S.} \bibnamefont{Centre}},
  \bibinfo{journal}{Journal of large-scale research facilities JLSRF}
  \textbf{\bibinfo{volume}{7}}, \bibinfo{pages}{A182} (\bibinfo{year}{2021}),
  ISSN \bibinfo{issn}{2364-091X},
  \urlprefix\url{https://www.journal-of-large-scale-research-facilities.org/index.php/lsf/article/view/182}.

\bibitem[{\citenamefont{Grimaldi et~al.}(2022)\citenamefont{Grimaldi,
  S\'anchez-Tejerina, Anjum~Aadit, Chiappini, Carpentieri, Camsari, and
  Finocchio}}]{PhysRevApplied.17.024052}
\bibinfo{author}{\bibfnamefont{A.}~\bibnamefont{Grimaldi}},
  \bibinfo{author}{\bibfnamefont{L.}~\bibnamefont{S\'anchez-Tejerina}},
  \bibinfo{author}{\bibfnamefont{N.}~\bibnamefont{Anjum~Aadit}},
  \bibinfo{author}{\bibfnamefont{S.}~\bibnamefont{Chiappini}},
  \bibinfo{author}{\bibfnamefont{M.}~\bibnamefont{Carpentieri}},
  \bibinfo{author}{\bibfnamefont{K.}~\bibnamefont{Camsari}}, \bibnamefont{and}
  \bibinfo{author}{\bibfnamefont{G.}~\bibnamefont{Finocchio}},
  \bibinfo{journal}{Phys. Rev. Appl.} \textbf{\bibinfo{volume}{17}},
  \bibinfo{pages}{024052} (\bibinfo{year}{2022}),
  \urlprefix\url{https://link.aps.org/doi/10.1103/PhysRevApplied.17.024052}.

\bibitem[{\citenamefont{Chowdhury et~al.}(2025)\citenamefont{Chowdhury, Aadit,
  Grimaldi, Raimondo, Raut, Lott, Mentink, Rams, Ricci-Tersenghi, Chiappini
  et~al.}}]{chowdhury2025pushingboundaryquantumadvantage}
\bibinfo{author}{\bibfnamefont{S.}~\bibnamefont{Chowdhury}},
  \bibinfo{author}{\bibfnamefont{N.~A.} \bibnamefont{Aadit}},
  \bibinfo{author}{\bibfnamefont{A.}~\bibnamefont{Grimaldi}},
  \bibinfo{author}{\bibfnamefont{E.}~\bibnamefont{Raimondo}},
  \bibinfo{author}{\bibfnamefont{A.}~\bibnamefont{Raut}},
  \bibinfo{author}{\bibfnamefont{P.~A.} \bibnamefont{Lott}},
  \bibinfo{author}{\bibfnamefont{J.~H.} \bibnamefont{Mentink}},
  \bibinfo{author}{\bibfnamefont{M.~M.} \bibnamefont{Rams}},
  \bibinfo{author}{\bibfnamefont{F.}~\bibnamefont{Ricci-Tersenghi}},
  \bibinfo{author}{\bibfnamefont{M.}~\bibnamefont{Chiappini}},
  \bibnamefont{et~al.}, \emph{\bibinfo{title}{Pushing the boundary of quantum
  advantage in hard combinatorial optimization with probabilistic computers}}
  (\bibinfo{year}{2025}), \eprint{2503.10302},
  \urlprefix\url{https://arxiv.org/abs/2503.10302}.

\end{thebibliography}
\end{document}